\documentclass[oneside]{article}
\usepackage[T1]{fontenc}
\usepackage{authblk}
\usepackage{float}
\usepackage{amsmath,bm}
\usepackage{graphicx}
\usepackage{xfrac}
\usepackage[english]{babel}
\usepackage{lmodern}
\usepackage[T1]{fontenc}
\usepackage{threeparttable}
\usepackage{multirow}
\usepackage{booktabs, caption, makecell}
\usepackage{mathtools}
\usepackage{graphicx}
\usepackage{esint}
\usepackage[unicode=true,
 bookmarks=false,
breaklinks=false,pdfborder={0 0 1},backref=false,colorlinks=false]
 {hyperref}
\usepackage[a4paper, total={210mm,297mm},margin=2.0cm]{geometry}

\usepackage{datetime}
\newdate{date}{22}{11}{2022}

\title{\textbf{A Review on Contact and Collision Methods for Multi-body Hydrodynamic problems in Complex Flows}}

\author[1]{S. Karimnejad}

\author[2]{A. Amiri Delouei\thanks{Electronic address: \texttt{a.amiri@ub.ac.ir; a.a.delouei@gmail.com}; Corresponding author}}

\author[3]{H. Ba\c sa\u gao\u glu}

\author[1]{M. Nazari}

\author[1]{M. Shahmardan}

\author[4,5]{G. Falcucci}

\author[3]{M.  Lauricella}

\author[3]{S. Succi}

\affil[1]{Faculty of Mechanical Engineering, Shahrood University of Technology, Shahrood, Iran.}

\affil[2]{Department of Mechanical Engineering, University of Bojnord, Bojnord, Iran.}

\affil[3]{Evolution Online LLC, San Antonio, TX 78292, USA.}

\affil[4]{Department of Enterprise Engineering “Mario Lucertini”, University of Rome “Tor Vergata”, Via del Politecnico 1, 00133 Rome, Italy.}

\affil[5]{Department of Physics, Harvard University, 29 Oxford Street, Cambridge, MA 02138, USA.}

\affil[6]{Istituto per le Applicazioni del Calcolo, Consiglio Nazionale delle Ricerche, Via dei Taurini 19, 00185, Rome, Italy.}

\affil[7,6]{Center for Life Nanoscience, Italian Institute of Technology, Viale Regina Margherita 295, 00161, Rome, Italy.}

\date{\displaydate{date}}

\begin{document}
%%%%% title : short title may not be used but TITLE is required.

%\title{A Review on Contact and Collision Methods for Multi-body Hydrodynamic problems in Complex Flows}
%%\title[short title]{A Review on Contact and Collision Methods}

%%%%%% author(s) : 

 %\author[Karimnejad]{S. Karimnejad\affil{1},
       %A. Amiri Delouei\affil{2}\comma\corrauth, and H. Ba\c sa\u gao\u glu\affil{3}, M. Nazari\affil{1}, M.  Shahmardan\affil{1}, G. Falcucci\affil{4}\comma\affil{5}, M.  Lauricella\affil{6}, and S. Succi\affil{6}\comma\affil{7}}  
 %\address{\affilnum{1}\ Faculty of Mechanical Engineering, Shahrood University of Technology, Shahrood, Iran. \\
           %\affilnum{2}\ Department of Mechanical Engineering, University of Bojnord, Bojnord, Iran. \\ 
           %\affilnum{3}\ Evolution Online LLC, San Antonio, TX 78292, USA. \\          
           %\affilnum{4}\ Department of Enterprise Engineering “Mario Lucertini”, University of Rome “Tor Vergata”, Via del Politecnico 1, 00133 Rome, Italy.\\
           %\affilnum{5}\ Department of Physics, Harvard University, 29 Oxford Street, Cambridge, MA 02138, USA.  \\
           %\affilnum{6}\ Istituto per le Applicazioni del Calcolo, Consiglio Nazionale delle Ricerche, Via dei Taurini 19, 00185, Rome, Italy. \\  
           %\affilnum{7}\ Center for Life Nanoscience, Italian Institute of Technology, Viale Regina Margherita 295, 00161, Rome, Italy.}
           
 %\corraddr{Department of Mechanical Engineering, University of Bojnord, Bojnord, Iran.
          %Email: a.amiri@ub.ac.ir; a.a.delouei@gmail.com}
\maketitle

%%%%% Begin Abstract %%%%%%%%%%%%%%%%%%%%%%%%%%%%%%%%%%%%%%%%%%%%%%%%%%%%%%%%%%%%
\begin{abstract}
Modeling and direct numerical simulation of particle-laden flows have a tremendous variety of applications in science and engineering across a vast spectrum of scales from pollution dispersion in the atmosphere, to fluidization in the combustion process, to aerosol deposition in spray medication, along with many others. Due to their strongly nonlinear and multiscale nature, the above complex phenomena still raise a very steep challenge to the most computational methods. In this review, we provide comprehensive coverage of multibody hydrodynamic (MBH) problems focusing on particulate suspensions in complex fluidic systems that have been simulated using hybrid Eulerian-Lagrangian particulate flow models. Among these hybrid models, the Immersed Boundary-Lattice Boltzmann Method (IB-LBM) provides mathematically simple and computationally-efficient algorithms for solid-fluid hydrodynamic interactions in MBH simulations. This paper elaborates on the mathematical framework, applicability, and limitations of various ‘simple to complex’ representations of close-contact interparticle interactions and collision methods, including short-range inter-particle and particle-wall steric interactions, spring and lubrication forces,  normal and oblique collisions, and mesoscale molecular models for deformable particle collisions based on hard-sphere and soft-sphere models in MBH models to simulate settling or flow of nonuniform particles of different geometric shapes and sizes in diverse fluidic systems.
\end{abstract}
%%%%% end %%%%%%%%%%%%%%%%%%%%%%%%%%%%%%%%%%%%%%%%%%%%%%%%%%%%%%%%%%%%%%%%%%%%%%%%%%%%

%%%%% Keywords %%%%%%%%%%%
%\keywords{Particulate flow, Collision, Close-contact Interaction, Immersed Boundary Method, Lattice Boltzmann Method.}

%%%% maketitle %%%%%%%%%%%%%%%%%%%%%%%%%%%%%%%%%%%%%%%%%%%%%%%%%%%%%%%%%%%%%%%%%%%%%%
%\maketitle

%%%% Start %%%%%%%%%%%%%%%%%%%%%%%%%%%%%%%%%%%%%%%%%%%%%%%%%%%%%%%%%%%%%%%%%%%%%%%%%%
\section{Introduction}
\label{sec1}
Theoretical physics has a long and distinguished tradition of grasping the essence of complex natural phenomena through the aid of relatively simpler models. Among others, remarkable examples in point include the Ising model for magnetism \cite{KM_SR16}, the Sandpile model for self-organized criticality \cite{M_PyhA91}, cellular automata for particle aggregation (DLA model), and lattice gas cellular automata for fluids {\cite{WG_2004}}. To a different extent, they all obey the overarching paradigm of ‘simple models for complex phenomena’, whereby complexity arises as to the emergent phenomena on top of repeated application of simple microscale rules. The strength of these models is their capability to capture universal features of the phenomena under inspection, regardless of specific details which may differ from system to system in the same so-called universality class.

Modern science and society, however, are increasingly confronted with phenomena whose complexity can no longer be quantitatively captured by the aforementioned ‘complex from simple’ paradigm, as clearly witnessed by the 2021 Nobel Prize in Physics.  The reason is that, due to strong nonlinear interactions extending over multiple decades in space and time, the divide between universality and specificity becomes hazy and dynamic, which means the usual separation between relevant and irrelevant degrees of freedom that powers most of modern theoretical physics is no longer viable. Under such circumstances, simple models must be supplemented with just the right amount of specificity, typically, but not necessarily, in the form of details belonging to a deeper (more microscopic) level of description. In the very act of integrating such details within a simple universal structure, these models lose their ‘simplicity’, opening up to a new paradigm that we may call ‘complex for complex’, i.e., complex models for complex phenomena. Modern computational physics offers several examples of this sort, and the subject of this review, particulate flows, inscribes precisely within such conceptual and computational framework.

Particulate suspension flow occurs in diverse scientific and technological applications. In the automotive industry, particle-fluid interactions and flow of particulate matter in thermal multiphase fluidic environments affect the design of combustion chambers and cyclone separators, as nano- to micron-size particles enhance the heat transfer in fluidic systems \cite{SCGF_PE_15, CR_CES_07, BDB_NC_15}. In industrial processes such as inkjet printing \cite{D_ARWMR_10, ZLC_PoF_20}, additive manufacturing of ceramics and cement \cite{NKI_CPBE_18}, and cosmetics \cite{HPF_ACIS_08}, the impact of a liquid drop containing solid particles on solid surfaces affects the deposition and splashing dynamics, in which particle-to-droplet size ratio and particle’s wettability play critical roles \cite{YC_IJHM_18}. In biomedical fields, the geometric shape, size, deformability, concentration, and surface properties are critical design parameters for engineered particles \cite{SFM_JFM_19} used for optimal deliveries of nanomedicine and chemotherapeutics to targeted tumor cells for enhanced treatments \cite{YTRD_JN_16, RJTK_NC_18, TWM_EODD_15, TDP_CTM_17, LKC_JN_19, MPM_Can_19}. In microfluidics, microfluidic devices are designed for enhanced shape- and size-based sorting, and enrichment of targeted cells (biotic particles such as rare circulating tumor cells and leukocytes) for non-invasive diagnosis or prognosis of chronic diseases \cite{PDM_CTC_17, GKK_MM_18, SRL_LoC_15}. In virus-induced cellular infection dynamics, virus (biotic particles) adsorption is associated with the frequency of their collisions with the cell surface, which is controlled by the nature of the cell, virus, and the ionic environment \cite{BB_Vir_14}. In energy sectors, sand erosion is a critical concern that can hamper the economic flow of hydrocarbon streams in oil and gas production from a reservoir of low formation strength \cite{NZI_Conf_18, SWB_Wear_15, ZMS_Wear_14}. Sand erosion, caused by the collision of fine sand particles (abiotic particles) with inner pipeline surfaces in oil and gas pipelines, often results in localized corrosion, material loss, failure in pipeline structural integrity, and production shutdown. Collisions induced by higher particle velocities in pipe systems result in higher erosion rates \cite{IF_Wear_14}. Biofilms formed by adhesion and subsequent colonization of free-swimming bacteria (planktonic chemotactic particles) on the inner walls of oil and gas pipelines could cause corrosion, which has been responsible for severe economic losses \cite{GKK_MM_18}. On the opposite end, free-swimming magneto-tactic bacteria (magnetic particles) in liquid media in a confined engineered domain were experimentally shown to generate low-voltage electricity through electromagnetic induction, which has the potential to be developed as an alternative bio-energy source \cite{BVJ_LAM_18}. In oil and gas exploration and developments in unconventional reservoirs, engineered nanoparticles (e.g., nanorobots) have become important tools for the analysis of the mineral composition, micropore structures, and rock physical properties \cite{HXB_PED_16}. In hydro-fracking, pressurized water mixed with proppant, including sand or ceramic particles, is pumped into a wellbore to hold newly created underground fractures open to allow the releases of natural gas from tight shale formations, which were previously unreachable reservoirs \cite{VBV_Sci_13, CFJ_PNAS_13, WCJR_RSER_14}. However, accidental spills or land application of hydro-fracking fluids often favor the transport of \textit{in-situ} colloidal particles and pollutants in soils, which could cause subsurface contamination. 

The main challenge in such diverse applications and scientific quest lies in thorough understanding and quantifying the underlying physical and mechanical processes governing particle-fluid and interparticle interaction dynamics and collision mechanisms in complex particle-laden flows \cite{YCF_JFM_21}. Complexity in such problems arises from the presence of hydrodynamically interacting and colliding particles dispersed in the form of mesoscale structures in a variety of fluid types such as Newtonian, viscoelastic, or stratified fluids in various flow conditions such as non-isothermal or turbulent flows. Although significant progress in experimental and numerical techniques has been accomplished in recent years and an open-source software has been developed for simulating colloidal systems \cite{BMTA_CPC_20}, studies in this field still need further advancements due to complexities in flow domain geometries, multi-body interactions of particles of nonuniform sizes and geometric shapes, the nature of the host fluid (e.g., non-Newtonian, viscoelastic, and compressible), and environmental conditions (e.g., thermal, turbulent, heterogeneous flow conditions), in addition to loosely constrained or unconstrained values of the interaction parameters used to describe interaction and collision dynamics of the particles. In this paper, we review various mathematically and physically ‘simple to complex’ close contact and collision dynamics models that have been recently implemented in MBH models to simulate particles flow and discuss their applicability and limitations in different flow settings. We also outline prospective developments for future applications.

%%%%%%%%%%%%%%%%%%%%%%%%%%%%%%%%%%%%%%%%%%%%%%%%%%%%%%%%%%%%%%%%%%%%%%%%%%%%%%%%%%%%%
\section{Numerical Methods for Particle-Fluid Hydrodynamics}
\label{sec2}
%%%%%%%%%%%%%%%%%%%%%%%%%%%%%%%%%%%%%%%%%%%%%%%%%%%%%%%%%%%%%%%%%%%%%%%%%%%%%%%%%%%%%

\begin{figure}[ht!]
\centering
\includegraphics[width=0.9\textwidth]{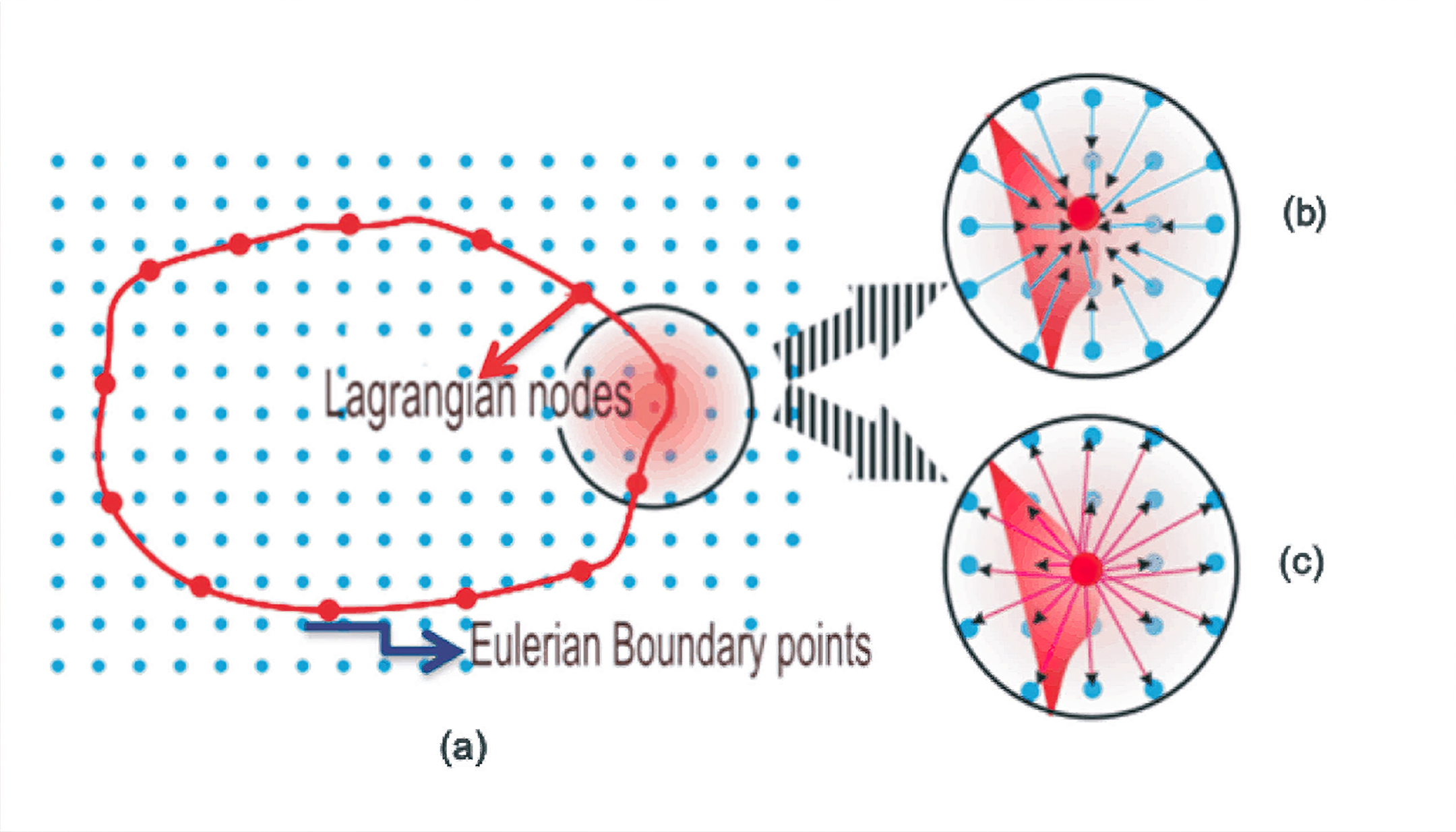}   
\caption{A schematic illustration of the IB-LBM. (a) An immersed body in the flow domain, whose boundary is described by a set of immersed boundaries (Lagrangian) nodes that do not conform to the Eulerian nodes, which are used to simulate the flow field, (b) momentum flux from the surrounding fluid into the immersed boundary nodes, (c) momentum flux from the immersed boundary nodes to the bulk fluid outside the immersed body (from \cite{DNKK_PhyA_16}  under conditions of Elsevier license N. 5355280607026). }
\label{fig:Fig1}
\end{figure}

Direct Numerical Simulations (DNS) are one of the commonly used MBH techniques to simulate hydrodynamic of particles. In the DNS, the fluid flow is described by the Navier-Stokes equations (NSE) in the Eulerian frame while the particle surface, represented by immersed (Lagrangian) boundary (IB) nodes, does not need to adhere to the Eulerian nodes \cite{DK_COCE_14, DK_CES_14}. Particle motion is described by the motion of immersed boundaries with tunable stiffness in response to local particle-fluid hydrodynamic forces in the vicinity of the particles \cite{DNK_CiCP_16, ET_IJHFF_16} and is computed by Newton’s laws of motion. A coupled Immersed Boundary-lattice Boltzmann method (IB-LBM) emerged as a promising numerical tool to simulate fluid-solid interaction (FSI) dynamics using moving boundaries and interfaces \cite{TLLG_ATE_20, MZY_JCP_20, WWQ_AMM_20, JLL_arch_20, WTL_JFS_20, DNKA_JAS_16, KDNS_ASME_19}. The IB-LBM has been successfully used in diverse computational fluid dynamics simulations, including incompressible viscous flow \cite{NSCP_PL_06}, multicomponent fluid flow \cite{LFD_JCP_16}, viscoelastic fluid flow \cite{SBSN_CCP_21}, compressible viscous flow \cite{DCB_Cong_21}, thermal flow \cite{JYHT_JCP_10}, and turbulent flow \cite{MF_PoF_21}, in addition to flow simulations with fluid-particle interactions \cite{FM_JCP_04}, including deformable liquid capsules \cite{ZJP_PB_07}, elastic filaments \cite{TLZL_JCP_11, AKDT_OE_22}, axisymmetric non-spherical particles \cite{CJXW_CEJ_21}, rigid spherical particles \cite{TMPC_CF_18}.  
 As shown in Fig. \ref{fig:Fig1}, the IB-LBM operates on two separate and independent computational domains, in which the fixed Eulerian nodes are used by the LBM to simulate the spatiotemporal evolution of the flow field, and the non-stationary Lagrangian nodes distributed along the immersed boundary are used by the IBM to simulate interparticle and particle-fluid hydrodynamics \cite{KKKS_LBM_17}. 

%%%%%%%%%%%%%%%%%%%%%%%%%%%%%%%%%%%%%%%%%%%%%%%%%%%%%%%%%%%%%%%%%%%%%%%%%%%%%%%%%%%%%%%%%%
%----- Figure 1 ----
%%%%%%%%%%%%%%%%%%%%%%%%%%%%%%%%%%%%%%%%%%%%%%%%%%%%%%%%%%%%%%%%%%%
%%%%%%%%%%%%%%%%%%%%%%%%%%%%%%%%%%%%%%%%%%%%%%%%%%%%%%%%%%%%%%%%%%%%%%%%%%%%%%%%%%%%%%%%%%

The IB-LBM has been improved on multiple fronts over the years. Many researchers developed new IB-LBM coupling schemes by imposing moving elastic or rigid immersed boundaries \cite{HW_PT_17}. When imposing elastic moving boundaries, boundary force is computed based on the boundary configuration governed by the physical law. An implicit IB scheme was proposed by Hao and Zhu \cite{HZ_CMwA_10} for 3D simulations of viscous flow past deformable sheets and flags. Liu et al. \cite{LGS_PoF_20} developed a coupled IBM-LBM with an absolute nodal coordinate formula and acceleration techniques to simulate large deformation in complex multiple flexible immersed structures. Guglietta et al. \cite{GBBF_SM_20, GBFS_SM_21} coupled the IB-LBM with the standard linear solid model to account for the red blood cell membrane viscosity. As for the rigid moving boundaries, researchers used various approaches to spread the IB force over fluid nodes by means of the Dirac delta function. Feng and Michaelides \cite{FM_JCP_04} proposed a penalty method to simulate 2D fluid-particle hydrodynamics. Their method allows the solid particles to deform slightly while a linear spring would restore the IB points back to their target location. A direct forcing immersed boundary method was introduced by Dupuis \cite{DCK_JPC_08} in which the IB forces were calculated by comparing the computed IB speed with the desired reference speed without applying the IB force.  Peng et al.\cite{PSCN_JCP_06} used a multi-block lattice in conjunction with the multi-relaxation-time LB and local grid refinement to enhance the numerical stability of the coupled IB-LBM. Xu et al. \cite{XTYL_JCP_18} developed an IB-LBM based on the dynamic geometry-adaptive Cartesian grid-based method to simulate solid-fluid interactions in moderate to high Reynolds number ($Re$) flows with enhanced numerical stability when the mass ratio of the immersed structure to the fluid is small.  

%%%%%%%%%%%%%%%%%%%%%%%%%%%%%%%%%%%%%%%%%%%%%%%%%%%%%%%%%%%%%%%%%%%%%%%%%%%%%%%%%%%%%
\subsection{Immersed Boundary Method (IBM) }
\label{sec2.1}
%%%%%%%%%%%%%%%%%%%%%%%%%%%%%%%%%%%%%%%%%%%%%%%%%%%%%%%%%%%%%%%%%%%%%%%%%%%%%%%%%%%%%

\begin{figure}[ht!]
\centering
\includegraphics[width=1.0\textwidth]{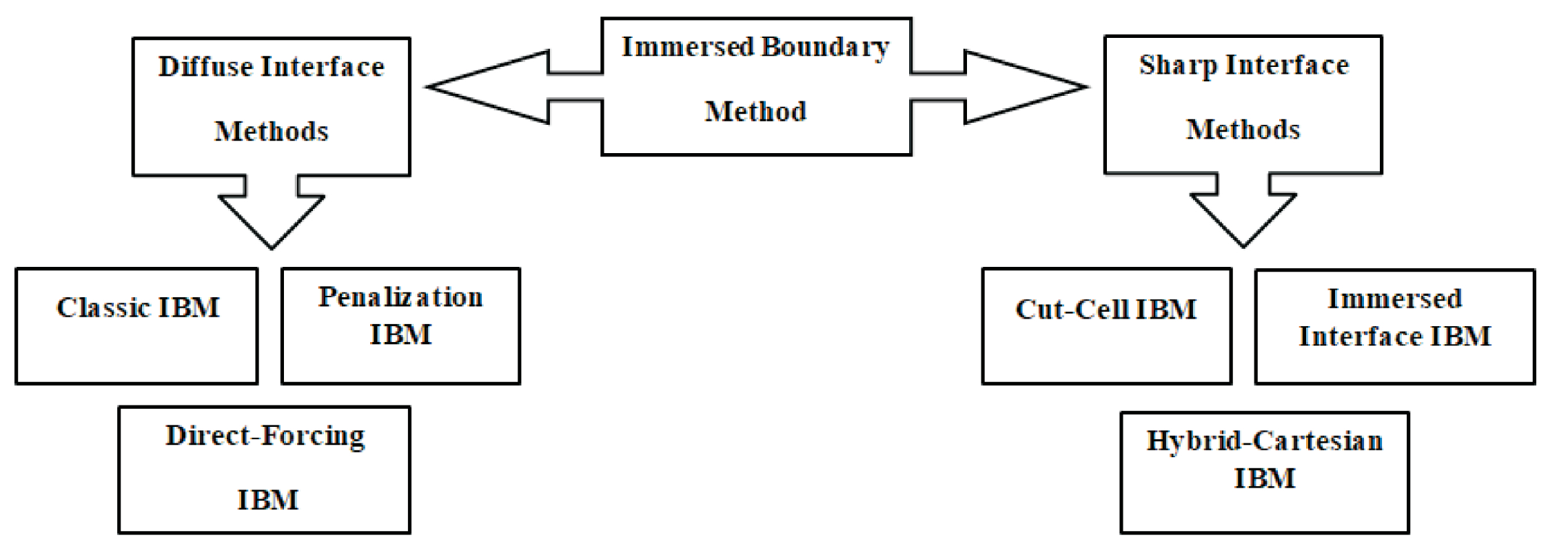}
\caption{The main variants of the immersed boundary method. }
\label{fig:Fig2}
\end{figure}

Body-fitted methods have been considered as a bottleneck in numerical simulations of particulate flows with multi-faceted geometries in a flowing fluid, as they typically require frequent remeshing, which could limit their use in modeling transient contacts and large deformations \cite{ZYTY_CMwA_16, JMS_CS_09}. As an alternative to body-fitted approach, IBM cropped up as a versatile numerical tool to simulate flow of particles with tunable deformability in complex fluidic systems. IBM was originally developed by Peskin \cite{CSP_JCP_72} in the 1970s to simulate blood flow in cardiac mechanics. Since then, it has been broadly used to simulate hydrodynamic interactions between deformable or rigid solid structures and the surrounding fluid in diverse applications \cite{SU_PAS_14, NYBS_SR_21}. In the IBM, the equations of fluid dynamics - represented by the NSE - is discretized in the Eulerian frame, whereas the elastic immersed structure is defined in the Lagrangian frame using a set of IB nodes. As the immersed solid structure deforms and moves, the IB nodes displace with respect to the underlying Eulerian mesh, and the interaction between the immersed structure and the fluid is accounted for by the FSI force, represented as a source term in the fluid momentum equation along with the external forces. In IBM simulations, the FSI term is determined by transforming the force density distributed on the immersed Lagrangian structure into body forces distributed onto the surrounding fluid nodes. In the disturbed fluid, the IB nodes move with the local fluid velocity when the no-slip boundary condition is imposed at the interface between the immersed structure and fluid. IBM can be divided into the diffuse interface scheme and the sharp interface scheme (Fig. \ref{fig:Fig2}), depending on how FSI is represented and simulated \cite{KKKS_LBM_17, CSP_AN_02, GP_ARFM_20, KC_JHFF_19}.

%%%%%%%%%%%%%%%%%%%%%%%%%%%%%%%%%%%%%%%%%%%%%%%%%%%%%%%%%%%%%%%%%%%%%%%%%%%%%%%%%%%%%%%%%%
%----- Figure 2 ----
%%%%%%%%%%%%%%%%%%%%%%%%%%%%%%%%%%%%%%%%%%%%%%%%%%%%%%%%%%%%%%%%%%%
%%%%%%%%%%%%%%%%%%%%%%%%%%%%%%%%%%%%%%%%%%%%%%%%%%%%%%%%%%%%%%%%%%%

%%%%%%%%%%%%%%%%%%%%%%%%%%%%%%%%%%%%%%%%%%%%%%%%%%%%%%%%%%%%%%%%%%%%%%%%%%%%%%%%%%%%%
\subsubsection{Diffuse Interface Methods }
\label{sec2.1.1}
%%%%%%%%%%%%%%%%%%%%%%%%%%%%%%%%%%%%%%%%%%%%%%%%%%%%%%%%%%%%%%%%%%%%%%%%%%%%%%%%%%%%%

In the diffuse interface scheme, forcing points at which the boundary force is evaluated are located on the immersed structure-fluid interface. The classical IBM (CIBM) \cite{CSP_JCP_72} is ideal for simulating flow of elastic immersed structures in a fluid. The IB nodes along the immersed boundary ($\Gamma$) are connected by either Stokes springs that obey Hook’s law or Maxwell elements that consist of a spring and a dashpot in series. The discrete delta functions (DDFs) are used to calculate hydrodynamic forces which are distributed from $\Gamma$ onto the adjacent Eulerian fluid nodes. Hydrodynamic forces between the immersed structure and the fluid ($f$) are calculated by

 \begin{equation}
f \left( \mathbf{x} \right) = \sum_{\mathbf{X \varepsilon \Gamma}} f \left( \mathbf{X} \right) \delta_h  \left( \mathbf{x} - \mathbf{X} \right)  \Delta X , \label{eq1}
 \end{equation}
 
 \noindent  where $\mathbf{X}$ is the position of the IB node and $\delta_h$ is the DDF. When the no-slip boundary condition is imposed at the immersed structure-fluid interface, the velocity at the IB node becomes equal to the local fluid velocity. The immersed boundary configuration can be used from the current time step (i.e., explicit approach) or previous time step (i.e., implicit approach) in hydrodynamic force calculations. The explicit approach is computationally inexpensive, but could suffer from numerical instabilities \cite{SW_JCP_99}, whereas the implicit approach is unconditionally stable, but computationally expensive \cite{MP_CMAME_08, TP_SIAM_92}. Lai and Peskin \cite{LP_JCP_20} developed a second-order accurate IBM that has less numerical viscosity than the first-order accurate IBM (the simplest explicit version), and hence, it emerged as a better simulator, especially in high Re flows. The CIBM \cite{CSP_JCP_72}, however, is not well-suited for simulating rigid immersed structures, as it would suffer from numerical instabilities due to the stiffness of the system. To overcome this problem, a direct-forcing immersed boundary method (DFIBM) was proposed \cite{FVO_JCP_20, JMY_ARB_97}, which temporally discretizes momentum equations in calculating hydrodynamic forces at $\Gamma$. If the Eulerian grid nodes coincide with the FSI, the time-discretized momentum balance equation is written as 
 
\begin{equation}
\frac{\mathbf{u}^{n+1} - \mathbf{u}^{n} }{\triangle t} = \left( \mathbf{u} \times \mathbf{\omega} \right) -\nabla p + \frac{1}{Re} \nabla^2 \mathbf{u} + \mathbf{f}, \label{eq2}
 \end{equation}
  
\noindent where $p$ is the pressure, $\mathbf{\omega}$ is the angular velocity, and $\mathbf{f}$ is the body force. The DDF of the DFIBM is used to exchange information between the immersed (Lagrangian) grid and Eulerian grid in the explicit and implicit methods \cite{KF_JCP_12, MU_JPC_05, SLL_CF_07, WZ_JCP_11}. Explicit time discretization in numerical simulations requires small time steps to maintain numerical stability. Uhlmann \cite{MU_JPC_05} proposed an explicit direct-forcing IBM in which forces at $\Gamma$ are computed by
 
  \begin{equation}
f \left( \mathbf{X_{\Gamma}} \right) =  \frac{\mathbf{u}^{n+1} \left( \mathbf{X}_{\Gamma} \right) -  \widetilde {\mathbf{u}} \left( \mathbf{X}_{\Gamma} \right) }{\triangle t}. \label{eq3}
 \end{equation}
 
 \noindent Su et al. \cite{SLL_CF_07} developed an implicit direct-forcing IBM to calculate forces at $\Gamma$
 
   \begin{equation}
\sum_{\mathbf{X \varepsilon \Gamma}} \left( \sum_{\mathbf{x \varepsilon \Omega }} \delta_h  \left( \mathbf{x} - \mathbf{X} \right)  \delta_h  \left( \mathbf{x} - \mathbf{X}_{\Gamma} \right) \triangle x \triangle X \right) f\left( \mathbf{X} \right)  =  \frac{\mathbf{u}^{n+1} \left( \mathbf{X}_{\Gamma} \right) -  \widetilde {\mathbf{u}} \left( \mathbf{X}_{\Gamma} \right) }{\triangle t}. \label{eq4}
 \end{equation}
 
\noindent As compared to its explicit counterpart, the implicit direct-forcing IBM yields higher accuracy in local velocities at the fluid-structure interface at the expense of higher computational costs that increase with the number of the IB nodes. However, if the elastic stiffness of the boundary does not dictate the timescale for the fluid motion, the implicit method could become computationally more efficient than the explicit method \cite{NFG_CMAME_08}. 

The DDF ($\delta_h$) can be envisioned as an inherent feature of the CIBM and DFIBM. A discretized smoothed approximation ($\phi$) to the DDF is used in calculating particle-fluid hydrodynamics at the IB nodes. Peskin \cite{CSP_AN_02} discussed the necessary postulates for extracting the desired interface function. $\delta_h$ is  a discrete delta function (with $h$ being the mesh width), which can be described in 2D as \cite{CSP_AN_02} 

   \begin{equation}
\delta_h \left( \mathbf{x} \right) = \frac{1}{h^2} \phi\left(\frac{x_1}{h}\right) \phi\left(\frac{x_2}{h}\right). \label{eq5}
 \end{equation}

The postulates discussed by Peskin \cite{CSP_AN_02} are imposed to hide the Eulerian computational lattice as much as possible. Continuity of $\phi$ is imposed in Eq. \ref{eq5} to avoid jumps in the velocity or applied force field as the Lagrangian nodes cross over the Eulerian grid plane. Several smoothed approximations (e.g., 2-points, 3-points, 4-points , and 4-point-$\cos$ functions) for $\phi$ have been proposed in the literature, as described below.

\vspace{0.3 cm}

\noindent 2-points function \cite{MU_JPC_05}:
 
\begin{equation}
 \phi\left(r \right) = \left\{ \begin{array}{rcl}
  1-|r|, & |r|\leq1, \\
  0, & otherwise,
\end{array}\right. \label{eq6}
 \end{equation}
 
  \noindent 3-points function \cite{SLL_CF_07}:
 
\begin{equation}
 \phi\left(r \right) = \left\{ \begin{array}{rcl}
 \frac{1}{6} \left( 5 - 3|r|-\sqrt{-3\left( 1- |r|\right)^2 + 1} \right), & 0.5 \leq|r| < 1.5, \\
 \frac{1}{3} \left( 1+ \sqrt{-3r^2 + 1 } \right), & |r| \leq 0.5, \\
  0, & otherwise,
\end{array}\right. \label{eq7}
 \end{equation}
 
 \noindent 4-points function \cite{ZYTY_CMwA_16}:
 
 \begin{equation}
 \phi\left(r \right) = \left\{ \begin{array}{rcl}
 \frac{1}{8} \left( 3 - 2|r|+\sqrt{1+4|r|-4r^2 } \right), & 0 \leq|r| < 1, \\
 \frac{1}{8} \left( 5 - 2|r| + \sqrt{-7 + 12|r| - 4r^2 } \right), & 1 \leq|r| < 2, \\
  0, & otherwise,
\end{array}\right. \label{eq8}
 \end{equation}
 
 \noindent 4-points-$\cos$. function \cite{WZ_JCP_11}:
 
\begin{equation}
 \phi\left(r \right) = \left\{ \begin{array}{rcl}
  \frac{1}{4} \left( 1+ \cos \left( \frac{\pi r}{2} \right) \right) , & |r|\leq 2, \\
  0, & otherwise,
\end{array}\right. \label{e9}
 \end{equation}
 
\begin{figure}[ht!]
\centering
\includegraphics[width=0.95\textwidth]{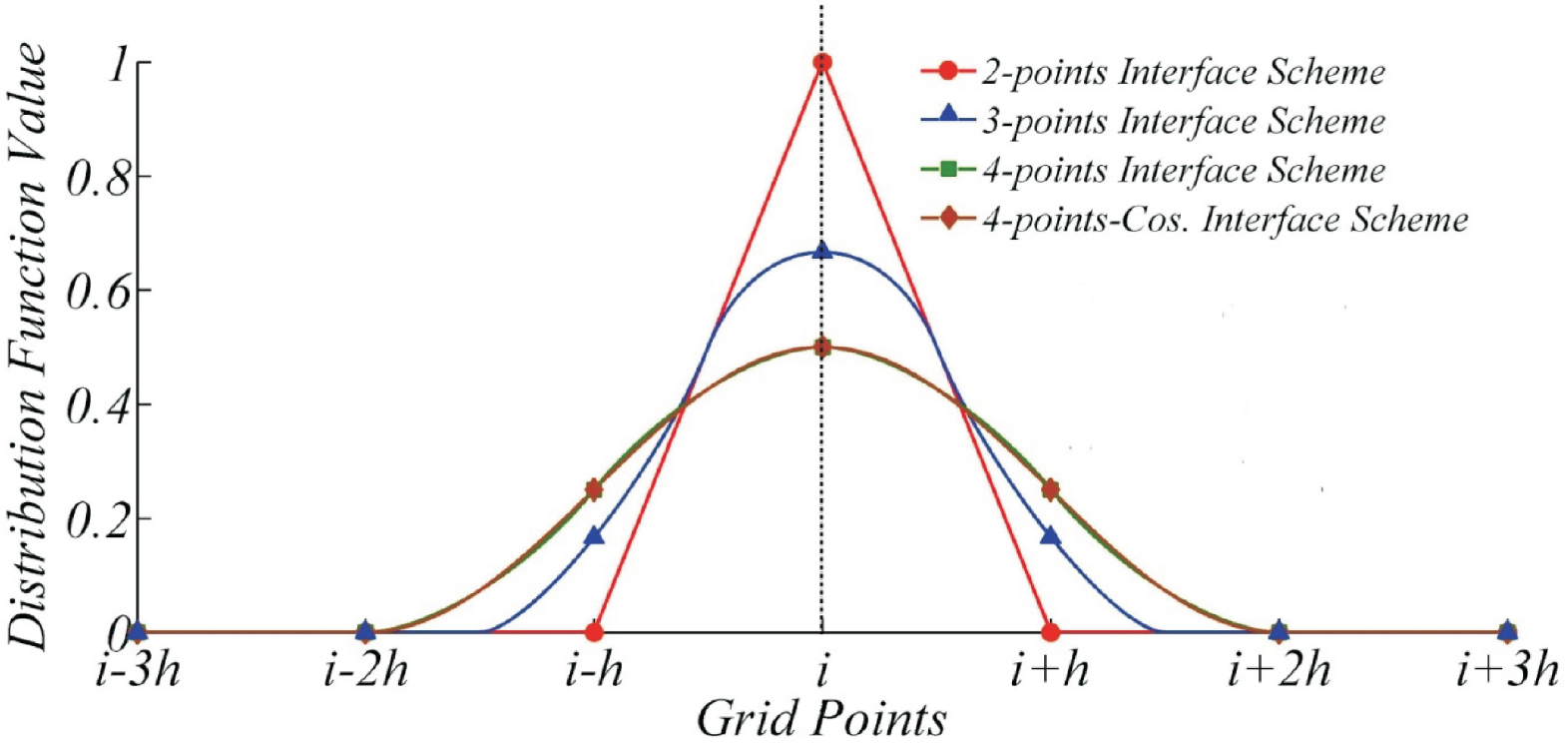}
\caption{Discretized smoothed approximation for $\phi$ (from \cite{DNK_CiCP_16} under Copyright © Global-Science Press 2015). }
\label{fig:Fig3}
\end{figure}

\noindent  Different distribution functions for $\phi$ near the boundary points are shown in Fig. \ref{fig:Fig3}.

%%%%%%%%%%%%%%%%%%%%%%%%%%%%%%%%%%%%%%%%%%%%%%%%%%%%%%%%%%%%%%%%%%%%%%%%%%%%%%%%%%%%%%%%%%
%----- Figure 3 ----
%%%%%%%%%%%%%%%%%%%%%%%%%%%%%%%%%%%%%%%%%%%%%%%%%%%%%%%%%%%%%%%%%%%
%%%%%%%%%%%%%%%%%%%%%%%%%%%%%%%%%%%%%%%%%%%%%%%%%%%%%%%%%%%%%%%%%%%

Because the no-slip boundary condition on the immersed structure surface is not satisfied exactly in the diffuse interface schemes, some streamlines may pass through the solid body \cite{SLC_JCP_07}. To overcome this problem, Shu et al. \cite{SLC_JCP_07}  proposed immersed boundary velocity correction method, in which the velocity field in the subsequent time step is predicted first by solving the NSE using the fluid velocities from the previous time-step while setting body forces to zero. The predicted new velocity field is then corrected to account for the effect of the body forces using, for example, the explicit Euler scheme. Thus, unlike the CIBM, the body forces are not pre-determined in this scheme. Subsequently, fluid velocities interpolated from the corrected velocities on the wall are set to wall velocity to enforce the no-slip boundary condition on the wall. 

In the penalization method (PM), the immersed solid structure is represented by a porous media with vanishing porosity at the solid wall \cite{AC_CRAcad_84, KS_arch_15}. As for the CIBM and DFIBM, the fictitious body force is smoothly smeared out across the immersed structure-fluid interface to ensure numerical stability. The hydrodynamic force between the immersed structure and fluid is computed by the penalized equation expressed as

\begin{equation}
 f\left( \mathbf{X} \right) = {\lambda_{x}}_{s} \left(  \mathbf{u^s} \left( \mathbf{X} \right)  -  \mathbf{u^f} \left( \mathbf{X} \right) \right),  \label{e10}
 \end{equation}

\noindent where $\lambda$ is the penalization parameter, and the superscripts $s$ and $f$ correspond to the solid and fluid phases. The choice of the penalization parameter value is crucial as large penalization parameter values could result in large stiffness. A binary masking function is commonly used to designate objects on a Cartesian grid by identifying if a particular computational element in the domain belongs to the solid phase or fluid phase. Penalized equations can be discretized independently of the geometry in the original problem. This leads to substantial reductions in solver time as it allows, for example, the use of simple Fast Fourier Transform (FFT)-based spectral solvers in Cartesian geometries. In addition, the penalized equations are solved without introducing additional discretization errors, and hence, the effects of numerical diffusion and dispersion that could occur low-order numerical schemes are avoided. Kim and Peskin \cite{KP_PoF_16, KP_PoF_07} used the penalty IBM to simulate elastic boundaries in fluid by implementing a massive boundary having all the mass of the elastic immersed boundary and a massless boundary. Kim and Peskin \cite{KP_PoF_16} extended the penalty IBM to simulate hydrodynamics interactions between a rigid body and the host fluid. 

%%%%%%%%%%%%%%%%%%%%%%%%%%%%%%%%%%%%%%%%%%%%%%%%%%%%%%%%%%%%%%%%%%%%%%%%%%%%%%%%%%%%%
\subsubsection{Sharp Interface Methods }
\label{sec2.1.2}
%%%%%%%%%%%%%%%%%%%%%%%%%%%%%%%%%%%%%%%%%%%%%%%%%%%%%%%%%%%%%%%%%%%%%%%%%%%%%%%%%%%%% 
 
 Sharp-interface IBMs modify the finite-difference stencils near the immersed boundary to ensure the accurate representation of the fluid boundary layer. In these methods, the immersed boundary is typically considered to be a boundary of the flow domain and the equations of fluid dynamics are solved on only one side of the interface. Sharp-interface IBMs were shown to reduce spurious-pressure oscillations in moving boundary problems, which is crucial when pressure field accuracy is a key concern \cite{CYH_PRE_13}.
Cut-cell IBM (CCIBM) was introduced by Clarke \cite{CSH_AIAA_86} for inviscid fluid flow problems, in which the Eulerian grid cells are cut by the boundaries of the solid structure. The grid cells are reconstructed along the immersed boundary, which results in locally unstructured and body-conforming meshes. Higher accuracy near the immersed boundary and inherent momentum conservation characteristics are the main advantages of the CCIBM \cite{MDH_JCP_10}. The method, however, is generally well-suited for 2D problems \cite{UMS_JCP_99, ICM_MCS_03, TD_MCS_20} as the extension of the cut-cells topology to 3D is tedious, yet doable \cite{SRD_JCP_20}. 

Leveque and Li \cite{LL_SIAM_94} proposed the Immersed Interface IBM (IIIBM) scheme for higher-order accurate FSI calculations. Similar to the CIBM scheme, the IIIBM calculates FSI forces to account for the effect of the immersed structure-fluid interface on the fluid domain but without using a DDF. Hence, the IIIBM can be viewed as the modified CIBM that enhances the accuracy around the boundaries of immersed structures \cite{GP_ARFM_20}. While the FSI force is distributed over the immersed boundary in the CIBM, jump conditions along fluid-immersed structure interfaces as a function of singular forces \cite{SX_AML_09} can be computed for elastic bodies based on constitutive relations \cite{XY_Diss_04} or using feedback approach for rigid bodies \cite{GP_ARFM_20, XW_JCP_06, TLK_JCP_09}. 

The Hybrid Cartesian IBM (HCIBM) \cite{JMY_CTR_97} can be viewed as an extension of the diffuse direct forcing method while maintaining the sharp interface representation of the immersed boundaries. Interpolation schemes \cite{TF_JCP_03, PDTM_IJNMF_06} are used to distribute FSI forces that are typically computed explicitly on the immersed boundaries and then transformed to the neighboring Eulerian grid. Various forms of HCIBM have been developed based on (i) how the velocities at the computational IB nodes are reconstructed \cite{IK_JCP_07}, (ii) whether the IB nodes coincide with the Eulerian grid nodes \cite{TF_JCP_03, MDBN_JCP_08}, and (iii) which type of the interpolation scheme is implemented \cite{PDTM_IJNMF_06, EB_CF_04, GS_JCP_05}. When HCIBM is implemented to simulate mobile immersed solid structures, non-physical force oscillations near the IB nodes could occur. Lee et al. \cite{LKCY_JCP_11} noted that such non-physical oscillations arise from spatial discontinuity in the pressure field and temporal discontinuity in the fluid velocity field. Refined grids near the boundary nodes, however, were shown to suppress such non-physical oscillations \cite{BGS_JCP_08}.

%%%%%%%%%%%%%%%%%%%%%%%%%%%%%%%%%%%%%%%%%%%%%%%%%%%%%%%%%%%%%%%%%%%%%%%%%%%%%%%%%%%%%
\subsection{Lattice Boltzmann Method (LBM) }
\label{sec2.2}
%%%%%%%%%%%%%%%%%%%%%%%%%%%%%%%%%%%%%%%%%%%%%%%%%%%%%%%%%%%%%%%%%%%%%%%%%%%%%%%%%%%%% 
 
The lattice Boltzmann method (LBM) has its root in kinetic theory and the cellular automaton concept and is ideal for mesoscale simulations that bridge the gap between microscale and macroscale processes \cite{SGRS_BAPS_20, SS_Book_01, HSB_ERL_89, CKOS_Sci_03, WGSR_MNRAS_20, BSV_PR_92}. The LBM has been used as a simulator in a broad range of applications including, for example, biophysics, computational aeroacoustics, fuel cell, heat transfer, phase change, complex states, natural convection, coarsening dynamics, and flow in granular porous media \cite{MS_JSP_02, DNKK_PhyA_16, CKMH_IJHMT_14, FCSS_EPL_08, BNSR_PRE_08, SDM_CiCP_21, MLTS_JFM_19, MF_Book_18, KF_Energ_18, GANF_Nat_21}. Its strong physical basis, explicit nature of calculations, and high level of scalability on parallel systems have played a crucial role in its rapid expansion and adaptation by the CFD community \cite{NBMC_MN_16, BASD_MN_13, Succi_Book_18}. In addition, the LBM does not generally require mesh quality check or remeshing in each time-step, and hence, can accommodate complex flow domain geometries without substantial computational overburden. Moreover, non-linear convective term in the NSE is replaced by a linear streaming term in the LBM, in which the streaming is literally exact and local mass and momentum conservation are accurate to machine round-off error  \cite{JDKG_JACM_20}. The accuracy of the streaming process near the solid objects, however, vary with the implementation of a first-order boundary condition or a second-order accurate boundary condition \cite{KKKS_LBM_17, OI_PoF_21}. In the LBM, the mesodynamics of the incompressible transient fluid flow can be described using a single relaxation time LBM (SRT-LBM), given by \cite{KKKS_LBM_17, SS_Book_01, QSO_Book_95, MPLS_PRE_15}

  \begin{equation}
 \label{e11} f_{i}\left(\mathbf{r+e}_{i}{\triangle t},t+{\triangle
 t}\right) -f_{i}\left(\mathbf{r},t \right) =\frac{\triangle t}{\tau} [
 {f_{i}^{eq}\left(\mathbf{r},t \right)-f_{i}\left(\mathbf{r},t
 \right) } ],
 \end{equation}
 
\noindent where $f_{i}$ is the complete set of population density of discrete velocities $\mathbf{e}_i$ at position $\mathbf{r}\left(x,y \right)$ and discrete time $t$ with a time increment of $\triangle t$. $\tau$ is the relaxation parameter defined as a function of kinematic viscosity of the fluid. For numerically stable solutions, $\tau > 0.5$. $f_{i}^{eq}$ is the discrete equilibrium Maxwell–Boltzmann distribution function in the $i^{th}$ direction approximated by the low Mach number mass and momentum conserving expansion \cite{KKKS_LBM_17, QSO_Book_95} 
 
  \begin{equation}
 \label{e12}f_{i}^{eq}=\omega_i \rho \left
 (1+\frac{\mathbf{e}_i\mathbf{\cdot}\mathbf{u}}{c_s^2}
 +\frac{(\mathbf{e}_i\mathbf{\cdot}\mathbf{u})^2}{2c_s^4}-
 \frac{\mathbf{u \mathbf{\cdot}u}}{2c_s^2}\right),
\end{equation}

\noindent where $\omega_i$ is the weight associated with $\mathbf{e}_i$ and $c_s$ is the speed of sound, $c_s= \triangle x /\sqrt(3) \triangle t$, $\rho$ is the local fluid density, and $\mathbf{u}$ is the local fluid velocity at the lattice node. The left-hand side of Eq. \ref{e12} describes the streaming of population densities, $f_{i}$, from a lattice node $\mathbf{r}$ to the closest neighboring lattice node in the direction of $\mathbf{e}_i$ on a regular lattice grid. The right-hand side of Eq. \ref{e12} describes the local collision process, in which population densities streamed into a particular lattice node $\mathbf{r}$ approach their equilibrium distribution upon collision at a rate of $1/\tau$. The local fluid density, $\rho$, and velocity, $\mathbf{u}$, at the lattice nodes are given by $\rho=\sum_{i} f_{i}$ and $\rho \mathbf{u}=\sum_{i} f_{i}\mathbf{e}_i+\tau \rho \mathbf{g}$, where $\mathbf{g}$ is the the acceleration due to an external force. Through the Chapman-Enskog approach, the LB method for a single-phase flow recovers the NSE within the limit of small Knudsen number for weakly compressible fluids \cite{MPLS_PRE_15, TS_EPL_05}. Different lattice geometries can be used in LBM simulations, such as D2Q9, D3Q15, D3Q19, and D3Q27, and $\mathbf{e}_i$ and $\omega_i$ vary with the lattice geometry. 

 To enhance the numerical stability of the LBM simulations especially at high $Re$ flow simulations, the multiple relaxation time (MRT) LBM was proposed \cite{LL_PRE_00}, in which $f$ is transformed into the velocity moment function and independent relaxation rate is used for each moment space from a diagonal relaxation matrix. In the MRT-LBM, only mass and flow momentum equations are conserved. The remaining moments are non-conserved and their equilibria are expressed in terms of conserved moments. The MRT-LBM was reported to be more stable than the SRT-LBM especially for low Mach and high $Re$ flow systems \cite{ATB_MM_14}. On the other hand, Nathen et al. \cite{NHKA_CF_18} reported that although the MRT-LBM did not show numerical instabilities at high $Re$ (e.g., $Re$=395 and 590), it resulted in unphysical spurious velocity oscillations in the bulk fluid, while the SRT-LBM yielded reasonable results. Although for the coupled IB-LBM, the LBM is typically used to simulate the flow field and the IBM is implemented to simulate the flow of particles with tunable stiffness, the LBM alone has been successfully used to simulate the settling and flow dynamics of rigid particles of different geometric shapes and sizes. In his pioneering work, Ladd \cite{AJL_JFM_94} extended the LB method to simulate the flow of finite-size, rigid spherical particles in a Newtonian fluid. Particle-fluid hydrodynamic calculations in the LBM rely on momentum exchanges between the particle in motion and the fluid in the immediate vicinity of the particle \cite{AJL_JFM_94, ALD_JFM_98}.

\begin{equation}
 \label{e13} \mathbf{F}_{\mathbf{r}_b}=-2\left[f^{\prime}_i\left(\mathbf{r}_v+\mathbf{e}_i \triangle t,t^{\ast}\right)+
\frac{\rho\omega_i}{c_s^2}\left(\mathbf{u}_{\mathbf{r}_b} \cdot
\mathbf{e}_i\right)\right]\mathbf{e}_i,
 \end{equation}

\noindent where $f^{\prime}_i$ is the population density in the $-\mathbf{e}_i$ direction at the post-collision time $t^{\ast}$, and $\mathbf{u}_{\mathbf{r}_b}$ is the local particle velocity at the boundary node $\mathbf{r}_b$. $\mathbf{r}_v$ and $\mathbf{r}_v+\mathbf{e}_i \triangle t$ are the intra-particle and extra-particle lattice nodes closest to the particle surface. The translational velocity, $\mathbf{U}_p$, and the angular velocity of the particle, $\Omega_p$, are computed in each time step by $\mathbf{U}_p\left( t+\triangle t\right) \cong \mathbf{U}_p\left(t\right) + \triangle t \left[ \frac{\mathbf{F_T \left( t \right) }}{{m_p}} +\frac{(\rho_p-\rho)}{\rho_p} \mathbf{g}  \right]$ and 
$\mathbf{\Omega}_p\left(t+\triangle t\right) \cong \mathbf{\Omega_p}\left(t\right)+\frac{\triangle t}{I_p}\mathbf{T}_T\left( t\right)$, where $\mathbf{F_T}$ and $\mathbf{T}_T$ are the total hydrodynamic force (including the particle–fluid hydrodynamic forces, in addition to interparticle and particle-wall interaction forces, which will be discussed in the following sections) and torque on the particle exerted by the surrounding fluid, respectively. $m_p$ is the particle mass, $I_p$ is the moment of inertia of the particle  and $\mathbf{u}_{\mathbf{r}_b}=\mathbf{U}_p+\mathbf{\Omega_p}\times\left({\mathbf{r}_b} 
- \mathbf{r}_c \right)$. The population densities at $\mathbf{r}_v$ and $\mathbf{r}_v+\mathbf{e}_i \triangle t$ are updated to account for momentum exchanges between the particle and bulk fluid in accordance with \cite{AJL_JFM_94}, 

\begin{equation}
 \label{e14} f^{\prime}_i\left(\mathbf{r}_v,t+\triangle t\right)=f_i(\mathbf{r}_v,t^{\ast})-\frac{2\rho \omega_i}{c_s^2}\left(\mathbf{u}_{\mathbf{r}_b} \cdot \mathbf{e}_i \right),  
 \end{equation}

\begin{equation}
 \label{e15}  f_i\left(\mathbf{r}_v+\mathbf{e}_i \triangle t,t+\triangle t\right)=f^{\prime}_i(\mathbf{r}_v+\mathbf{e}_i \triangle t,t^{\ast})+\frac{2\rho
\omega_i}{c_s^2}\left(\mathbf{u}_{\mathbf{r}_b}  \cdot \mathbf{e}_i \right).
 \end{equation}
 
 The LBM formulation above has been commonly used to simulate flow of circular-cylindrical particles in Newtonian fluid flow. The extension of the model to simulate settling or flow trajectories of 2D angular-shaped (e.g., boomerang-, star-, rectangular-, hexagonal-shaped particles) and circular-shaped (e.g., elliptical) particles in Newtonian or non-Newtonian fluid flow were provided by Ba\c sa\u gao\u glu et al. \cite{BBSF_MN_19, BSWB_SR_18}. In the coupled IBM-LBM, the IBM is typically used to calculate the FSI forces, and LBM is used to simulate the flow field by incorporating the FSI as the source term in the LB equation. Regardless of which model to use, interparticle hydrodynamic interactions, close contacts and collisions are pivotal in particle flow simulations.   

 %%%%%%%%%%%%%%%%%%%%%%%%%%%%%%%%%%%%%%%%%%%%%%%%%%%%%%%%%%%%%%%%%%%%%%%%%%%%%%%%%%%%%
\section{Collision Models Based on Repulsive and Lubrication Forces }
\label{sec3}
%%%%%%%%%%%%%%%%%%%%%%%%%%%%%%%%%%%%%%%%%%%%%%%%%%%%%%%%%%%%%%%%%%%%%%%%%%%%%%%%%%%%% 
 The likelihood of near contact interactions and interparticle collisions depend on various features such as particle motion, particle properties, fluid parameters, and particles volume fractions. For example, Ernst and Sommerfeld \cite{ES_JFE_12} showed that the effect of the particle volume fraction on the particles’ collision rate is less significant for the particles with small Stokes numbers than the particles with large Stokes numbers where the collision rate of the particles is strongly related to the adopted volume fraction. Different volume fraction thresholds were used in the literature to differentiate dilute and dense suspensions regimes. For example, the volume fraction $\leq 2\%$ was considered to form a dilute regime while $\geq 10\%$  was considered to form a dense regime \cite{SB_TC_20}.  In other studies, the dilute regime was defined by the volume fraction of $\leq 6\%$ \cite{MPWG_CPS_03} and  semi-dilute regime was defined by the volume fraction of $10-15\%$ \cite{GP_JFM_18}. 
 
 %%%%%%%%%%%%%%%%%%%%%%%%%%%%%%%%%%%%%%%%%%%%%%%%%%%%%%%%%%%%%%%%%%%%%%%%%%%%%%%%%%%%%
\subsection{Drafting-Kissing-Tumbling – Benchmark Problems }
\label{sec3.1}
%%%%%%%%%%%%%%%%%%%%%%%%%%%%%%%%%%%%%%%%%%%%%%%%%%%%%%%%%%%%%%%%%%%%%%%%%%%%%%%%%%%%% 
 
\begin{figure}[ht!]
\centering
\includegraphics[width=0.85\textwidth]{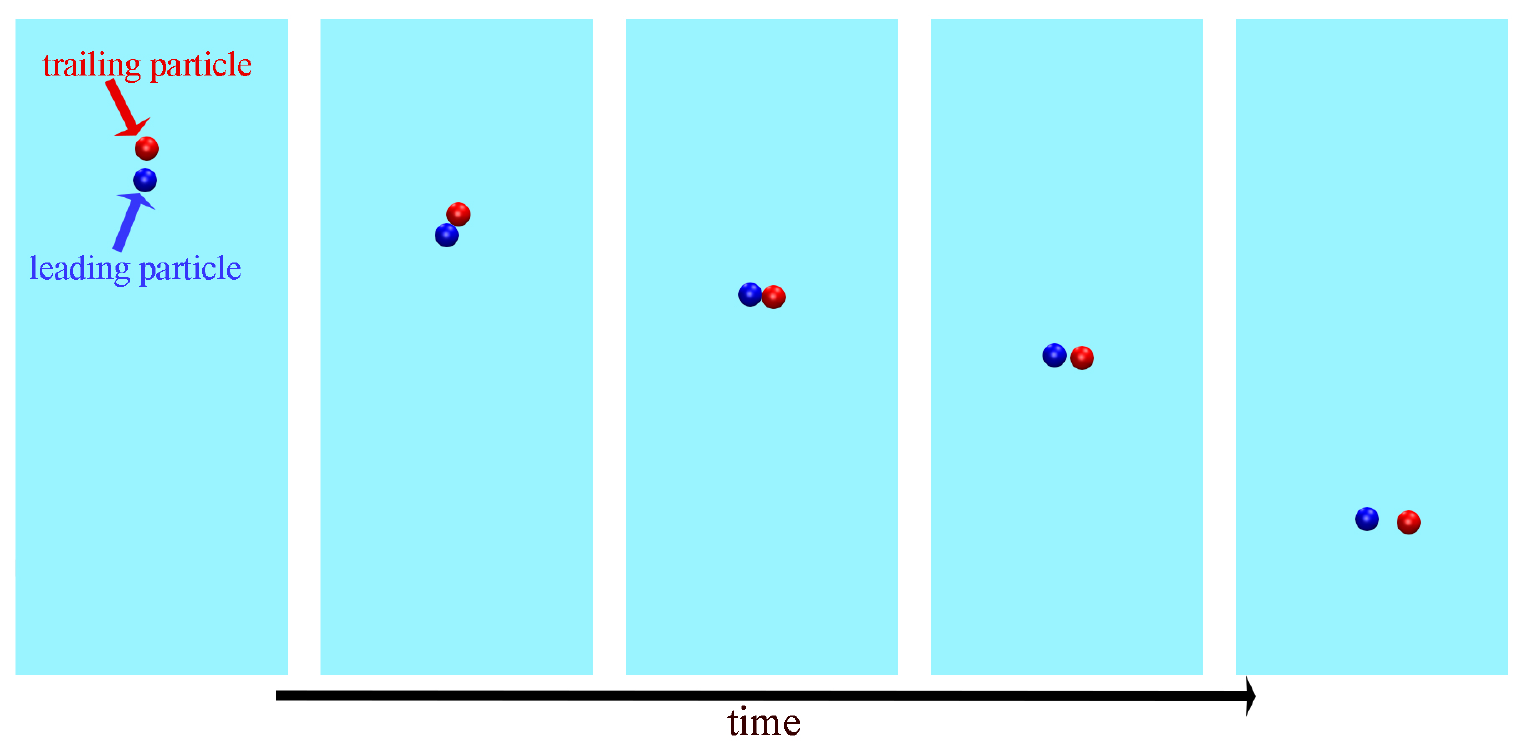}
\caption{Sketch showing gravity-driven settling of two spherical particles that exhibit the DKT behavior in an initially quiescent fluid. }
\label{fig:Fig4}
\end{figure}

\begin{figure}[ht!]
\centering
\includegraphics[width=1.0\textwidth]{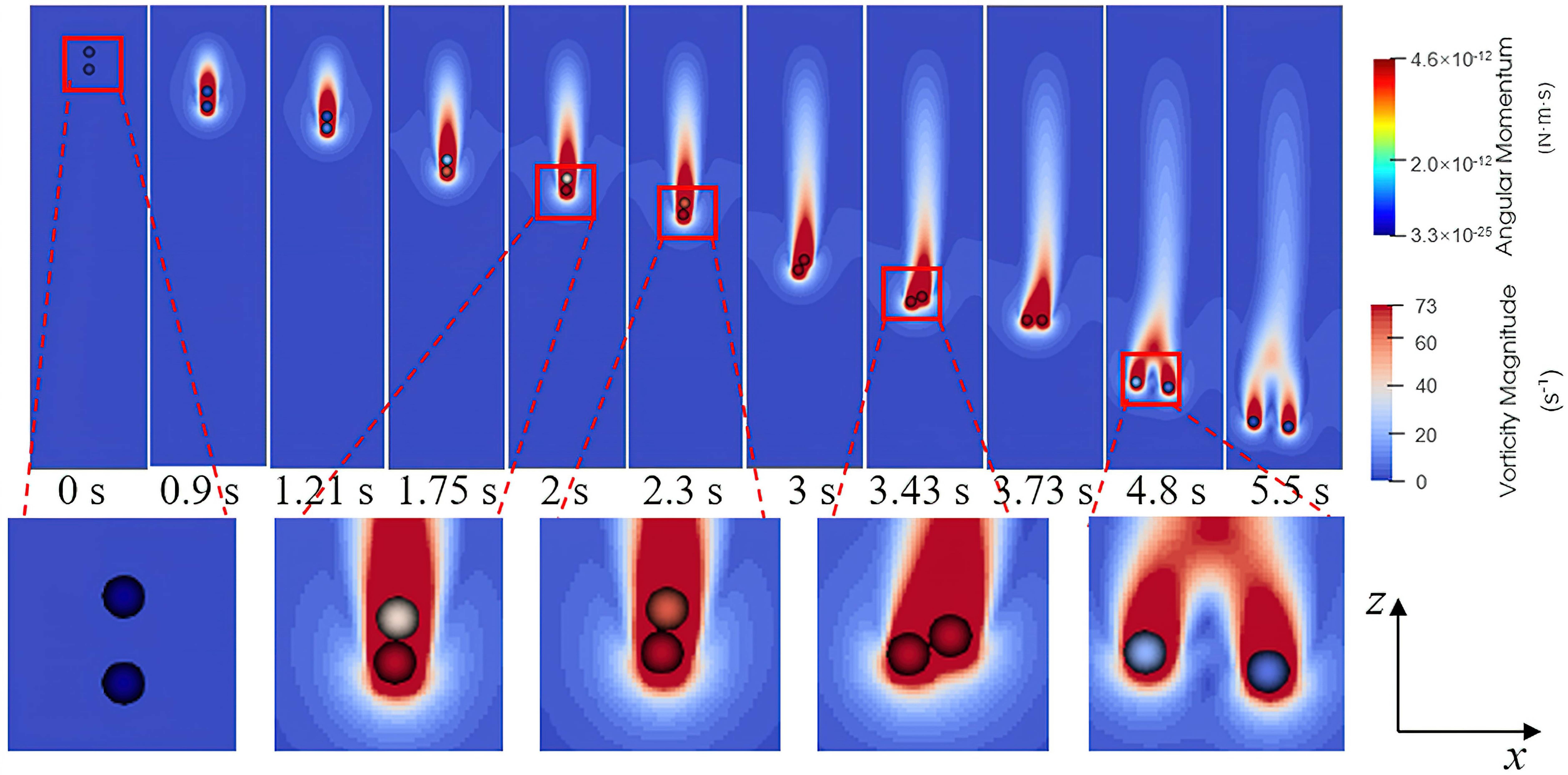}
%\caption{Vorticity contours during the sedimentation of two circular particles at (a) $t$=1.5 s; (b) $t$=1.8 s; (c) $t$=2.5 s; (d) $t$=3.5 s (adapted from \cite{WGM_CF_14}). }
\caption{Distribution of the magnitudes of fluid vorticity and particle angular momentum in the sedimentation process two in-tandem particles with terminal Re = 18 (from \cite{li2022ibm} under conditions of the Creative Commons Attribution CC BY license).}
\label{fig:Fig5}
\end{figure}

 Hydrodynamic interactions among mobile particles play a crucial role in their settling and flow dynamics and thermodynamics. Trustworthy calculations of particulate flow at the particle scale are essential for realistic representation and quantification of collision processes \cite{CLCZ_PT_21}. At the benchmark scale, two particles denser than the bulk fluid are shown to display a Drafting, Kissing and Tumbling (DKT) behavior as they settle in an initially quiescent fluid, if they are initially positioned with a small diagonal gap in their settling direction \cite{JFLS_PhFL_87}. The DKT of two spherical particles typically involves three distinct intra-particle interaction modes. At early times, the leading particle produces a weak pressure wake, which pulls the trailing particle toward itself, known as the Drafting stage. The Drafting stage is followed by the Kissing stage, in which the trailing particle moves into the close vicinity of the leading particles, which subsequently contact and collide with the leading particle. In the Tumbling stage, the particles tumble and separate from each other due to wake’s inertial influence. The DKT behavior of two particles was experimentally captured using a high-speed camera by Diaz-Goano \cite{DGMN_JCP_03}. Fig. \ref{fig:Fig4} shows the typical DKT behavior of the particles as they settle in an initially quiescent fluid due to gravity.  Vortex shedding of the settling particles during the DKT process is demonstrated in Fig. \ref{fig:Fig5} \cite{li2022ibm}.
 
%%%%%%%%%%%%%%%%%%%%%%%%%%%%%%%%%%%%%%%%%%%%%%%%%%%%%%%%%%%%%%%%%%%%%%%%%%%%%%%%%%%%%%%%%%
%----- Figure 4 ----
%%%%%%%%%%%%%%%%%%%%%%%%%%%%%%%%%%%%%%%%%%%%%%%%%%%%%%%%%%%%%%%%%%%
%%%%%%%%%%%%%%%%%%%%%%%%%%%%%%%%%%%%%%%%%%%%%%%%%%%%%%%%%%%%%%%%%%%

Wang et al. \cite{WGM_CF_14} analyzed the effect of the size ratio of the circular-cylindrical particles (circular particles hereafter) of the same mass density and their initial separation distance on their DKT behavior in a Newtonian fluid.  The authors demonstrated that the same-size particles displayed repetitive DKT behavior. When different size particles were used, their DKT behavior was sensitive to their initial positions. For example, when the smaller particle was placed below the larger one in the settling direction, their initial separation distance did not affect the overall DKT behavior due to the faster settling of the trailing particle. When the size ratio of the particles increased, the particles exhibited the DKT behavior at earlier times, but the repetitive DKT behavior vanished as the trailing particle swapped its position with the leading particle. Conversely, when the smaller particle was initially placed above the larger particle in the settling direction, their initial separation distance positively correlated with delays in their DKT behavior. 

%%%%%%%%%%%%%%%%%%%%%%%%%%%%%%%%%%%%%%%%%%%%%%%%%%%%%%%%%%%%%%%%%%%%%%%%%%%%%%%%%%%%%%%%%%
%----- Figure 5 ----
%%%%%%%%%%%%%%%%%%%%%%%%%%%%%%%%%%%%%%%%%%%%%%%%%%%%%%%%%%%%%%%%%%%
%%%%%%%%%%%%%%%%%%%%%%%%%%%%%%%%%%%%%%%%%%%%%%%%%%%%%%%%%%%%%%%%%%%%%%%%%%%%%%%%%%%%%%%%%%

\begin{figure}[ht!]
\centering
\includegraphics[width=0.7\textwidth]{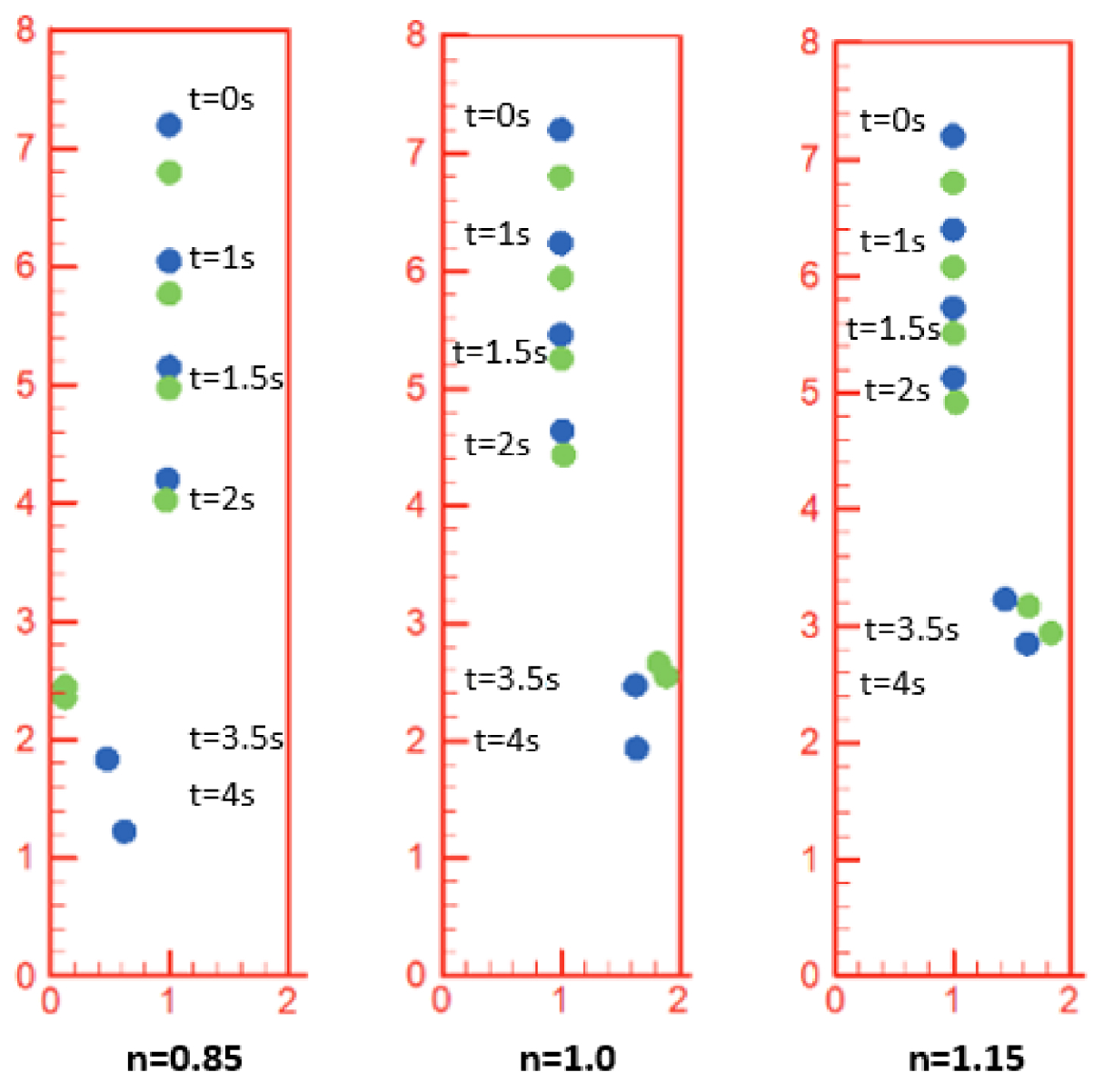}
\caption{Sedimentation of two circular particles in a bounded domain at different times in a shear-thinning ($n$=0.85), Newtonian (n=1), and shear-thickening ($n$=1.15) fluid flow at $t$=0, 1, 1.5, 2, 3.5, 4 s (from \cite{DNKK_PhyA_16} under conditions of Elsevier license N. 5355280607026). }
\label{fig:Fig6}
\end{figure}

The fluid type was also shown to influence the DKT behavior of the particles. Amiri Delouei et al. \cite{DNKK_PhyA_16} simulated the DKT phenomena of two circular particles in non-Newtonian fluid, represented by a power-law fluid (Fig. \ref{fig:Fig6}). Simulations revealed that the elapsed time of the Drafting and Kissing phases were shorter in the shear-thinning fluid than in the shear-thickening fluid. Moreover, a smaller non-Newtonian fluid index of the power-law fluid resulted in increases in the amplitude of the angular velocity of the particles.

%%%%%%%%%%%%%%%%%%%%%%%%%%%%%%%%%%%%%%%%%%%%%%%%%%%%%%%%%%%%%%%%%%%%%%%%%%%%%%%%%%%%%%%%%%
%----- Figure 6 ----
%%%%%%%%%%%%%%%%%%%%%%%%%%%%%%%%%%%%%%%%%%%%%%%%%%%%%%%%%%%%%%%%%%%
%%%%%%%%%%%%%%%%%%%%%%%%%%%%%%%%%%%%%%%%%%%%%%%%%%%%%%%%%%%%%%%%%%%%%%%%%%%%%%%%%%%%%%%%%%
 
 The DKT behavior displayed greater variability in more complex settings. Settling of two disk-shaped particles in a viscoelastic fluid was numerically shown to display either periodic interactions in the DKT sequence or chain formations, depending on the magnitude of the elasticity number of the viscoelastic fluid and the Mach number \cite{PG_Arch_16}. In heterogeneous fluids, the buoyancy-induced vortical structures and stratified jets behind the particles were shown to govern the dynamics of the particle pair interaction in stratified fluids \cite{DA_PRE_13}. Even in the case of weak stratification, the DKT occurred with a prolonged Kissing time while the rate of change in the orientation of the particles decreased. When two identical particles settled under gravity in a confined domain with finite internal heat sources, the Drafting sequence and interparticle interactions were faster, if the heat source was small, whereas the Kissing and Tumbling sequence did not occur at larger internal heat sources due to development of natural convection currents around the particles \cite{SS_PE_15}. Moreover, the relative temperature of two settling particles was shown to alter their DKT behavior at the different Grashof numbers ($Gr$) \cite{YCL_IJHMT_17}. For example, when both particles were hot, hot upward stream induced by the particles delayed the start of the DKT behavior but prolonged the DKT period. At $Gr$ =2,000, the DKT process vanished and hot upward stream led to abnormal oscillations in the settling trajectory of the trailing particle. When both particles were cold, repulsive interactions between the particles were strongest at $Gr$=1,000. When one particle was hot and the other one was cold, vortex shedding and stream oscillations behind the cold particle resulted in abnormal oscillations in velocities of both particles.
 
\begin{figure}[ht!]
\centering
\includegraphics[width=1.\textwidth]{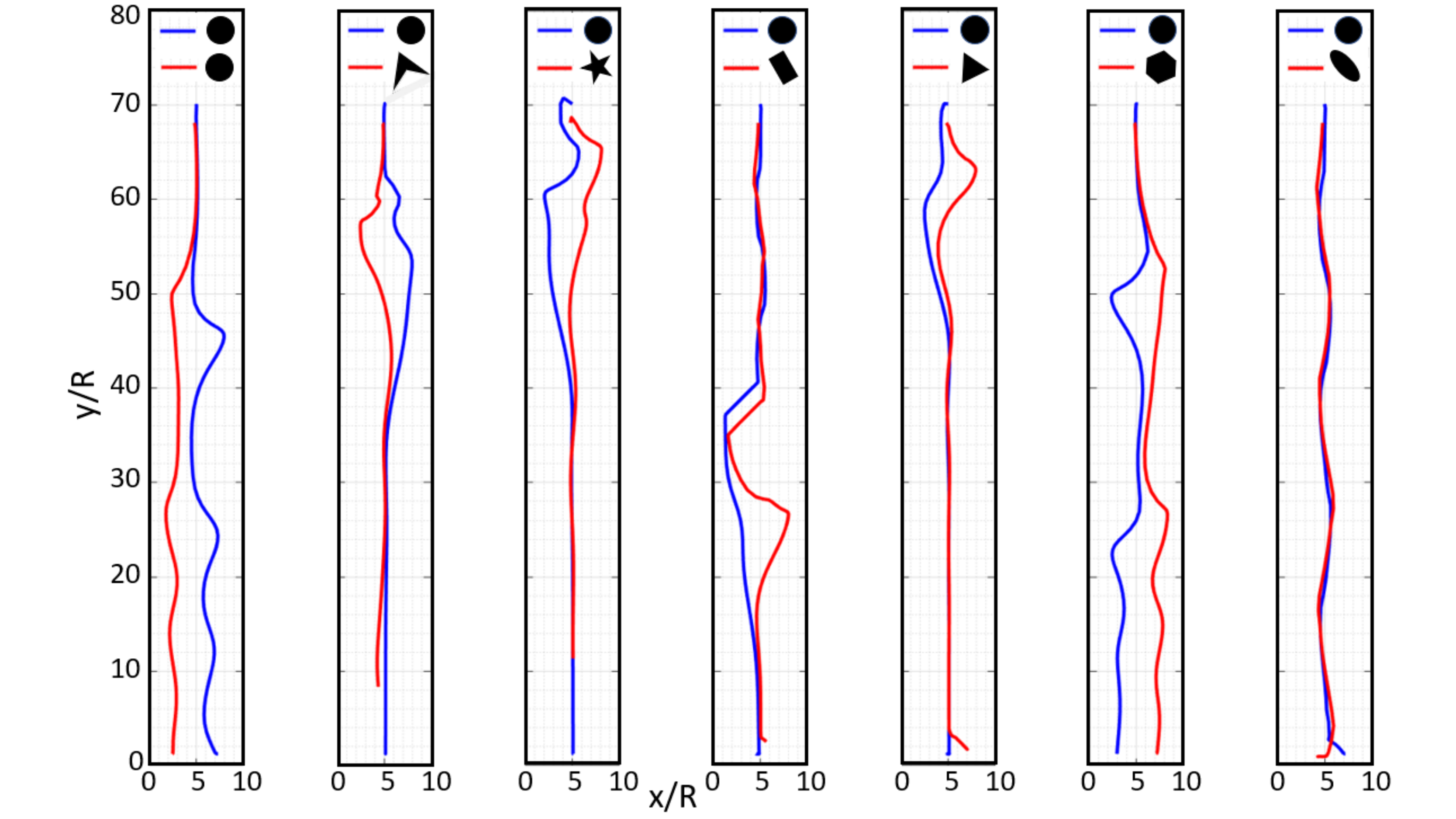}
\caption{2D LBM simulation of DKT trajectories of particles with different geometric shapes in an initially quiescent Newtonian fluid in a bounded domain, using the LBM by Ba\c sa\u gao\u glu et al. \cite{BSWB_SR_18}. The initially trailing particle was circular in shape in all simulations, but the shape of the leading particle varied in each scenario. The size of the trailing circular particle was $R=w /10$, here $w$ is the width of the domain perpendicular to the settling direction. Lennard-Jones potentials (discussed in Section \ref{sec3.6}) were used to simulate interparticle and particle-wall interactions. Simulations ended one of the particles reached the bottom boundary. The initial release location of the trailing particle was $(W/2,L-10R)$, where $L$ is the length of the domain $(L=8W=80R)$. The initial position of the leading particle was  $(W/2–R/3, L-12R)$. %Both particles were 5\% denser than the fluid. 
Among all the pairs, initially leading boomerang- and star-shaped particles considerably lagged behind the initially trailing circular particle when they approached the bottom boundary. }
\label{fig:Fig7}
\end{figure}

 In light of the numerical and experimental analyses of gravity-driven settling of two particles discussed above, near-contact interactions and collisions are inevitable in suspended particle flow \cite{FM_JCP_04, BSWB_SR_18}, which are influenced by the size, shape, temperature, and initial position of the particles and non-Newtonian, isothermal, and heterogeneous nature of the fluidic environment. Numerical simulations of near-contact interactions and collisions of mobile particles are often computationally demanding \cite{FM_JCP_04, BSWB_SR_18}. Sophisticated particle flow models for complex MBH problems involving near-contact interaction and collision dynamics of swarm of particles in complex fluidic environments should be able to accurately simulate the benchmark DKT analyses discussed above. The DKT analyses have been commonly performed using curved-shaped (e.g., spherical and ellipsoidal) particles. Wang et al. \cite{WFQZ_PT_21} simulated the DKT behavior of two identical circular (disk)-shaped particles, whose boundaries were represented by 16- or 8-vertices and analyzed the effect of polygonal representation of particles’ surface on their DKT.  Although the authors noted that there was no such benchmark for the coupling between polygon/polyhedron and fluid, LBM simulations involving settling or flow trajectories of polygon/polyhedron particles were already presented in \cite{BSWB_SR_18}. Our 2D LBM analysis (based on the LBM formulation in \cite{BSWB_SR_18}) revealed that the DKT trajectories of the particles of nonidentical geometric shapes are sensitive to their shapes (Fig. \ref{fig:Fig7}). In simulations presented in Fig. \ref{fig:Fig7}, the channel length $L$ =3.08 cm, the channel width $W$=0.38 cm, gravitational field strength $g$ = 9.81 cm/s$^2$, fluid kinematic viscosity $\nu$=0.01 cm$^2$/s, relaxation parameter $\tau$=0.6, grid spacing $\triangle x$=3.85$\times$10$^{-3}$ cm, and temporal spacing $\triangle t$=4.93$\times$10$^{-5}$ s. The surface area of all the particles was set to 4.65$\times$10$^{-3}$ cm$^2$, and they were all 5\% denser than the fluid. The radius of the circular particle was 3.85$\times$10$^{-2}$ cm. The major and minor axes of the elliptical particles were 1.09$\times$10$^{-1}$ cm and 5.44$\times$10$^{-2}$ cm.  The inner and outer angles of the boomerang-shaped particle were 60$^\circ$ and 30$^\circ$. The length of the rectangular particle is 1.05$\times$10$^{-1}$ cm. The initial tilting angle of the boomerang-, elliptical-, star- , triangle-, rectangle-, and hexagonal- shaped particles were randomly set to 60$^\circ$, 180$^\circ$, 300$^\circ$, 30$^\circ$, 90$^\circ$, and 0$^\circ$ to demonstrate the ability of the LBM to accommodate different initial placements of different-shaped particles in simulations. Geometrical representation of these particle shapes, calculation of the position of their vertices based on the geometric information provided above, and their geometric shape-specific mass and inertia calculations are provided in \cite{BSWB_SR_18}.
 
 %%%%%%%%%%%%%%%%%%%%%%%%%%%%%%%%%%%%%%%%%%%%%%%%%%%%%%%%%%%%%%%%%%%%%%%%%%%%%%%%%%%%%%%%%%
%%%%%%%%%%%%%%%%%%%%%%%%%%%%%%%%%%%%%%%%%%%%%%%%%%%%%%%%%%%%%%%%%%%%%%%%%%%%%%%%%%%%%%%%%%
%----- Figure 7 ----
%%%%%%%%%%%%%%%%%%%%%%%%%%%%%%%%%%%%%%%%%%%%%%%%%%%%%%%%%%%%%%%%%%%

%%%%%%%%%%%%%%%%%%%%%%%%%%%%%%%%%%%%%%%%%%%%%%%%%%%%%%%%%%%%%%%%%%%%%%%%%%%%%%%%%%%%%
\subsection{Repulsive Force Models}
\label{sec3.2}
%%%%%%%%%%%%%%%%%%%%%%%%%%%%%%%%%%%%%%%%%%%%%%%%%%%%%%%%%%%%%%%%%%%%%%%%%%%%%%%%%%%%% 

As the particles approach each other in a fluid, the viscous force is exerted on both particles by the fluid squeezed in the narrow gap between their solid surfaces. If the particle’s momentum exceeds the steric interaction strength between the solid surfaces, they can come into close or even direct contact (e.g., dry collision). At this stage, the particles may undergo either inelastic or elastic collisions, through which they can either repel or adhere to each other \cite{GPHJ_JCP_01}. Although such collisions happen in reality, they are commonly overlooked due to the prevailing repulsive forces between solid surfaces over a finite time interval \cite{GPHJ_JCP_01, GPHJ_CMAME_00, DT_IJNMF_06}. However, because of numerical errors or in an attempt to avoid truncation errors in numerical simulations, collisions of the particles are unavoidable \cite{GPHJ_JCP_01}. Collisions are considered to be smooth if the contact velocities of the solid surfaces match during collisions \cite{GPHJ_JCP_01, GPHJ_CMAME_00}, which are easier to simulate numerically. Such smooth collisions can occur, for example, during the settling of rigid particles in a confined domain filled with a viscous fluid. 
Ladd \cite{AJL_PRL_96} noted that computational methods (e.g., DNS methods on a fixed grid) would require at least one lattice node between solid surfaces in lubrication force calculations. Hofler and Schwarzer \cite{HS_PRE_00} noted that when the separation distance between solid surfaces of particles in close contact is less than one lattice unit, lubrication forces could be under-estimated. Computationally efficient algorithms are often required to accurately resolve such short-range lubrication forces and the viscous stress between adjacent particles. The lubrication force acts as a repulsive force on the particles and its strength varies with the size, shape, and surface roughness of the particles, fluid characteristics (e.g., viscosity, temperature, heterogeneity), and the flow regime. Thin film flow between the solid surfaces in close contact may need to be treated at a different spatial scale in numerical simulations depending on its spatial extent. A finer mesh may be needed to resolve the hydrodynamics over the thin film accurately at the expense of high computational costs \cite{HS_PRE_00, AW_CF_09}.

In the collision-stage, the particles come into direct contact, at which they can either repel or adhere to each other. Various numerical approaches have been proposed to simulate the collision of particles \cite{BBGE_PoF_13, JT_CMAME_96, MS_IJMF_92}. Johnson and Tezduyar \cite{JT_CMAME_96} developed a numerical method to study 3D particle-fluid interactions, involving 100 particles. To simulate the flow and collision of the particles, they placed a small fictitious domain around the particles that can deform as the particles approach each other. They considered the normal collision forces but neglected the tangential collision forces at the collision stage. Their method allowed 1-2 collisions in each time step, as compared to only one collision throughout the entire simulation time in its earlier counterparts. Ardekani and Rangel \cite{AR_JFM_08} proposed the use of a contact force as a gauge to predict when the collision ends, in which the contact force was derived from the conservation of linear momentum. In their model, the collision begins when the particles touch or the separation distance between them becomes less than or equal to the magnitude of the surface roughness and ends when the normal collision force becomes non-positive. The surface roughness was presumed to increase with the increased number of collisions but remains unchanged during post-collisions. Because velocities of the particles are not explicitly updated, the method is prone to numerical instabilities. Sommerfeld \cite{MS_IJMF_01} developed a stochastic collision model based on the kinetic theory and generation of fictitious particles with the given size and velocity, whereby no information was required on the actual position and direction of motion of the surrounding real particles. The fictitious particles represented the local particle-phase properties. The collision probability built on the kinetic theory determines if the collision to occur. For the model validation, the authors used the results from large eddy simulations (LES) involving uniform-size or binary-size particles dispersed in homogeneous isotropic turbulent flow. A comparison of collision frequencies of uniform-size or binary-size particles in a homogeneous isotropic turbulent flow condition simulated by the LES and Sommerfeld’s model exhibited a close agreement. The repulsive force models (RFMs), including lubrication force, spring force, and steric interaction force models, are typically used in numerical simulations to prevent particle-particle and particle-wall overlaps after the particles come into close contact with each other or stationary objects. Some other RFMs, however, allow the particles to move into close or direct contact, or even to slightly overlap. In this case, the RFM can be computed based on body-body and body-wall collision forces. The lubrication models typically accommodate the effect of the trapped thin fluid flow between solid surfaces \cite{HW_PT_17}. The repulsive force strength varies with the stiffness or magnitude of the repulsive strength ($\varepsilon$), and the separation distance between the particles or between the particle and the immobile solid surface ($\triangle x$).

\begin{figure}[ht!]
\centering
\includegraphics[width=0.55\textwidth]{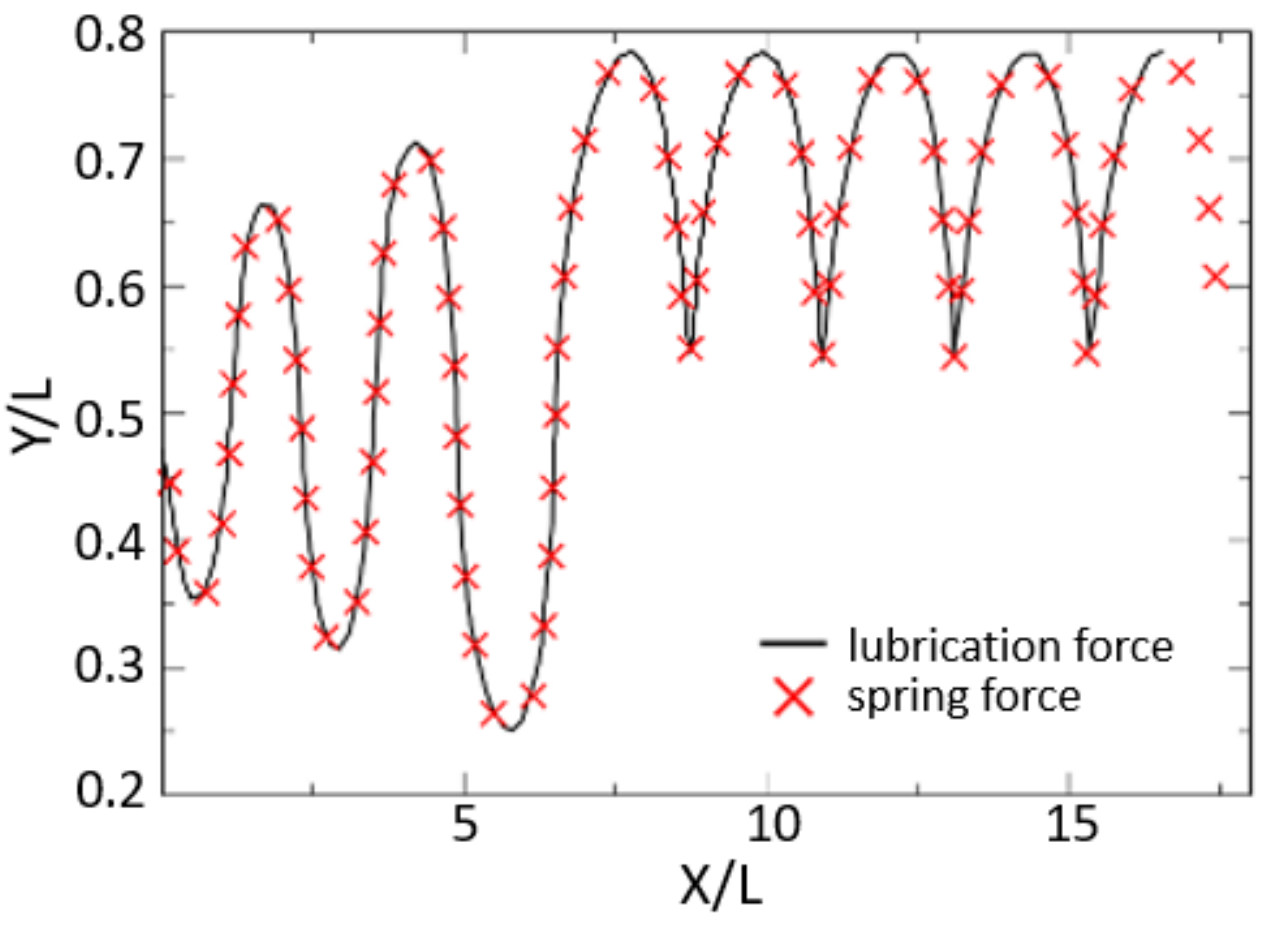}
\caption{Comparison of simulated settling trajectories of an elliptical particle in a narrow domain, in which particle-wall interactions were simulated using the lubrication force correction or the spring force correction. The simulated trajectories were identical in these simulations, in which the aspect ratio, blockage ratio, and the density ratio were set to 2.0, 16/13, and 1.1, respectively (digitized reconstruction from \cite{XCRY_JFM_09}). }
\label{fig:Fig8}
\end{figure}

The repulsive force has normal and tangential components. The normal component acts in the direction from the center of a particle to the center of a neighboring particle. Hence, the torque on the particle arising from the normal component of the repulsive force vanishes for the circular or spherical particles, but not for non-circular/non-spherical particles \cite{XCRY_JFM_09, FYZ_Parti_17}. Two broadly used RFMs for rigid particles involve the spring force \cite{GPHJ_JCP_01, GPHJ_CMAME_00} and lubrication force \cite{YB_JCP_94} models, which were shown to provide consistent results in particle flow simulations \cite{XCRY_JFM_09} (Fig. \ref{fig:Fig8}). In some cases, the distinction between the lubrication and spring force models is vague from a mathematical standpoint, and hence, they can be used interchangeably. Hard-sphere or soft-sphere approaches \cite{CB_JFM_85, CS_Geotech_79} are also used for simulating particle flow in granular systems. Other RFMs based on, for example, the Lennard-Jones potentials have also been widely used to simulate short-range interactions \cite {YP_JCIS_04}.

%%%%%%%%%%%%%%%%%%%%%%%%%%%%%%%%%%%%%%%%%%%%%%%%%%%%%%%%%%%%%%%%%%
%----- Figure 8 ----
%%%%%%%%%%%%%%%%%%%%%%%%%%%%%%%%%%%%%%%%%%%%%%%%%%%%%%%%%%%%%%%%%%%

%%%%%%%%%%%%%%%%%%%%%%%%%%%%%%%%%%%%%%%%%%%%%%%%%%%%%%%%%%%%%%%%%%%%%%%%%%%%%%%%%%%%

The RFMs are easier to formulate and apply to curved-surfaced particles than arbitrary-shaped particles in close contact \cite{LV_JSP_01, ACBB_IJMF_16, YDCM_PT_16}. For example, the surface-to-surface distance between two spherical particles can be computed from their radius of curvatures by \cite{LCP_SIAM_02}

\begin{equation}
 \label{e16}  x = c_1 + \frac{R_1}{\|c_1-c_2 \|} \left( c_2 - c_1 \right),
 \end{equation}

\begin{equation}
 \label{e17}  y = c_2 + \frac{R_1}{\|c_1-c_2 \|} \left( c_1 - c_2 \right),
 \end{equation}
 
 \noindent where $\left( x,y \right)$ are the position of the nearest boundary nodes of two adjacent spherical particles, $c_1$ and $c_2$ are the centroid of the particles, and $R_1$ and $R_2$ are the radius of the particles. Surface-to-surface distance calculations for non-spherical particles are computationally more involved. Lin and Han \cite{LCP_SIAM_02} proposed an iterative scheme to calculate the closest boundary points of two ellipsoidal particles in close contact. The algorithm is outlined in Table \ref{ClosestBnd} \cite{LCP_SIAM_02, ARJ_JFM_07}. 
 
%%%%%%%%%%%%%%%%%%%%%%%%%%%%%%%%%%%%%%%%%%%%%%%%%%%%%%%%%%%%%%%%%%%%%%%%%%%%%%%%%%%%%%%%%
%---------------------------------------------------------------------
\begin{table}[! htbp]
\centering 
\caption{The algorithm to calculate the closest boundary nodes of two ellipsoids in close contact.}
\small
\tabcolsep=0.11cm
\begin{threeparttable}
\begin{tabular}{l c c }
\toprule
\thead{ Steps } 
 &\thead{ Explanation} 
 &\thead{} 
 \\ \midrule

Step 1  &  \thead{Choose two boundary points $(x,y)$, shown by black circles, on the surface of \\ ellipsoids ($E_1$ and $E_2$  )} &  \\
Step 2  &  \thead{Insert two spheres inside being tangential to the inner surfaces at $(x,y)$}  &  \\
Step 3  & \thead{Check if the line segment connecting the centers of the two 
spheres is \\ entirely contained in $E_1 \cup E_2$. If it does, then the distance   
equals zero, \\ and the particles are tangent to each other (the closest possible  position). \\ Otherwise, a new (x,y) is chosen}  &  \\
Step 4  &  \thead{Choose new (x,y)  at the intersections of the line segments and the \\ boundaries as depicted in the figure}  & \\
Step 5  &  Repeat steps 2 to 4 until it converges  &  \\
  & {\includegraphics[width=0.4\textwidth]{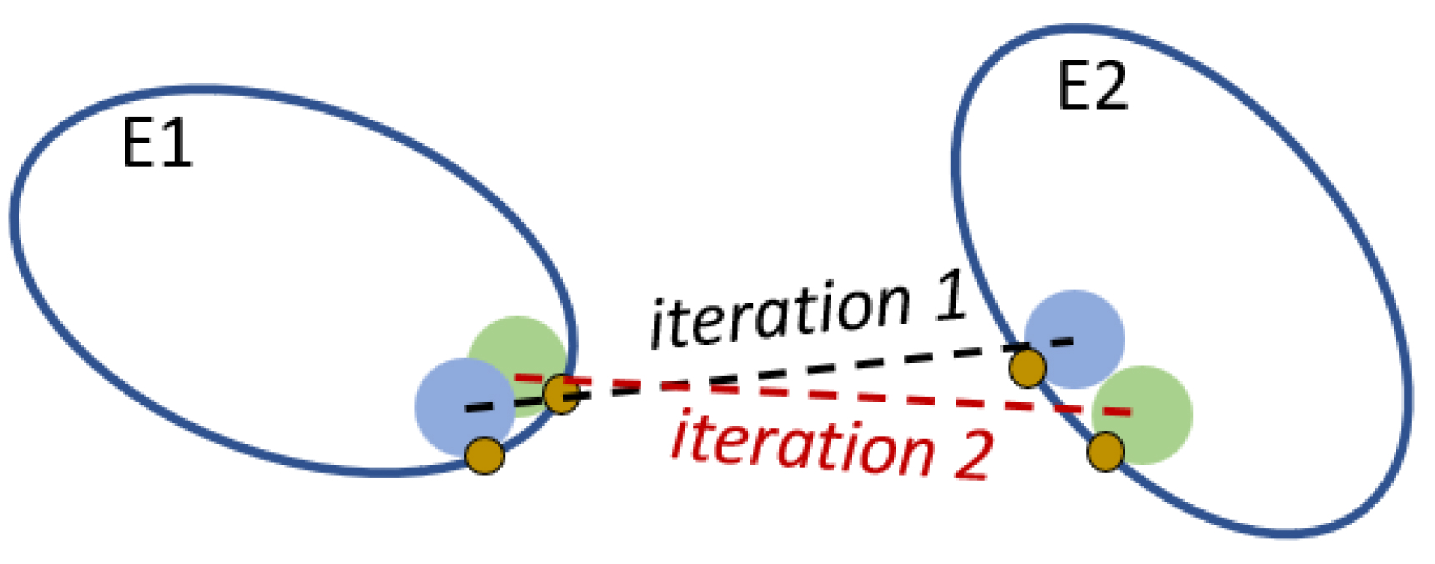}}   &  \\

\bottomrule\addlinespace[1ex]
\end{tabular}

\end{threeparttable}
\label{ClosestBnd}
\end{table}

%%%%%%%%%%%%%%%%%%%%%%%%%%%%%%%%%%%%%%%%%%%%%%%%%%%%%%%%%%%%%%%%%%%%%%%%%%%%
 
 Computational time in particle flow simulations was shown to increase significantly with smaller gaps between the particles and longer threshold interaction distances in the RFMs \cite{JT_CMAME_96}. Although $\varepsilon$ is a critical parameter in RFMs, there is no unified approach to determine its value. Therefore, alternative RFM formulations independent of $\varepsilon$ have been also proposed. The threshold interaction distance ($\zeta$) is another critical RFM parameter. 

Using the LBM, Ten Cate et al. \cite{CNDV_PoF_02} simulated trajectories and settling velocities of a spherical particle obtained from the particle imaging velocimetry (PIV). In agreement with the earlier findings by Ladd \cite{AJL_PRL_96}, Ten Cate et al. \cite{CNDV_PoF_02} noted that in the absence of $\zeta$, the contact moment between solid surfaces in near contact was delayed excessively due to the lubrication force. Mountrakis et al. \cite{MLH_PRE_17} investigated the sensitivity of interparticle interactions to the relaxation time, mesh size, and interpolation kernel through 3D IB-LBM simulations. Unlike in \cite{GPHJ_JCP_01, GPHJ_CMAME_00, DT_IJNMF_06}, the authors introduced a correction for the lubrication force to facilitate head-on collision of the particles while maintaining the interparticle interaction force in the particle center-to-particle center direction. The authors studied temporal changes in the center-to-center interparticle separation distance ($\triangle x$) and the separation distance at which lubrication force failed $\left( \triangle x_{fail} \right)$ \cite{MLH_PRE_17}, and concluded that large-size particles reached $\triangle x_{fail}$ faster than the small-size particles. Unexpectedly, $\triangle x_{fail}$ was not correlated to the relaxation time of LBM, and hence, to the fluid viscosity \cite{MLMA_JCS_15}. 

When direct contacts or overlaps between the particles and/or the particle and the wall occur, the total repulsive force can be computed by summing up the Body-Body Collisions (BBC) force and Body-Wall Collisions (BWC) force. The resultant total force ($\mathbf{F}_t^{Col}$) and torque ($\mathbf{T}_t^{Col}$) can be computed as \cite{GPHJ_JCP_01, GPHJ_CMAME_00, DT_IJNMF_06} 

 \begin{equation}
 \label{e18}  \mathbf{F}_t^{Col} = \sum _{|\triangle \mathbf{x}_{pp}| \leq \zeta} \mathbf{F}_{R_f}^{pp} + \sum _{|\triangle \mathbf{x}_{pw}| \leq \zeta} \mathbf{F}_{R_f}^{pw},
 \end{equation}
 
 \begin{equation}
 \label{e19}  \mathbf{T}_t^{Col} = \sum _{|\triangle \mathbf{x}_{pp}| \leq \zeta} \triangle \mathbf{x}_{pp} \times \mathbf{F}_{R_f}^{pp} + \sum _{|\triangle \mathbf{x}_{pw}| \leq \zeta} \triangle \mathbf{x}_{pw} \times \mathbf{F}_{R_f}^{pw},
 \end{equation}
 
 \noindent where $R_f$ denotes the repulsive force. The subscripts $p$ and $w$ refer to the particle and wall. The repulsive force and torque vanish when $|\triangle \mathbf{x}|>\zeta$.  If the separation distance between the particles or between the particle and the wall is less than $|\triangle \mathbf{x}|$, the RFM concludes that the collision is to occur. Therefore, the BBC and BWC forces are computed and their contributions are added to the total forces and torques on the particles in near contact. The total forces and torques are then used to update the translation and angular velocities, and the new position of the particles. 
 
 %%%%%%%%%%%%%%%%%%%%%%%%%%%%%%%%%%%%%%%%%%%%%%%%%%%%%%%%%%%%%%%%%%%%%%%%%%%%%%%%%%%%%
\subsection{Lubrication Force Models}
\label{sec3.3}
%%%%%%%%%%%%%%%%%%%%%%%%%%%%%%%%%%%%%%%%%%%%%%%%%%%%%%%%%%%%%%%%%%%%%%%%%%%%%%%%%%%%% 
 
 Interparticle and particle-wall interaction, close-contact, and collision dynamics are complex processes, which have been simulated using mathematically and physically relatively simpler to more complex lubrication force (LF) models. A relatively simple LF model \cite{CNDV_PoF_02} (LF-1) was formulated to correct the velocity of a settling spherical particle in a confined container when the separation distance ($|\triangle \mathbf{x}|$) between the particle and wall becomes less than a grid size. The LF-1 was defined based on the particle size ($R_p$), particle velocity normal to the wall surface ($\mathbf{U}_  \perp$), and dynamic viscosity of the fluid ($\eta$). It relies on a well-defined threshold interaction distance ($\zeta$) constrained to a single lattice unit, but does not involve the stiffness parameter ($\varepsilon$) to adjust the repulsive interaction strength. When the LF-1 was used with the LBM presented in \cite{DV_AICHEJ_99, ES_IJHFF_95}, the resultant MBH model simulated the experimentally observed settling of a spherical particle successfully at $Re$=1.5 and $Re$=31.9 \cite{CNDV_PoF_02}. The LF-1 can potentially be used to simulate settling dynamics of multiple, equal-sized spherical particles in Newtonian or non-Newtonian fluids. The performance of the LF-1 to simulate flow dynamics of a larger number of particles at higher $Re$ flows, however, has not been reported. Another relatively simple LF model (LF-2) was formulated based on the cluster implicit method \cite{NL_PRE_02} to calculate LF among different-sized spherical particles in near contact in a cluster as a function of their separation distance and relative velocities, and $\eta$ \cite{AJL_PRL_96, NL_PRE_02, AL_PRE_02}. The LF-2 can be implemented in MBH simulations with Newtonian or non-Newtonian fluids. It relies on $\zeta$ (not well-constrained), but not on $\varepsilon$. The model was reported to provide accurate results only at very small separation distances (e.g., $\triangle x \sim 0.01 R_p$). In addition, MBH simulations with the LF-2 resulted in significant error ($\sim 30\%$) in the total torque computed on a spherical particle rotating in a low viscosity fluid \cite{NL_PRE_02}. Thus, these relatively simpler LF models are expected to have limited applications in practice. 
 
 Singh et al. \cite{SHJ_IJMFF_03} developed a mathematically simple, but physically more realistic LF model (LF-3) that uses an elastic model to simulate the LF between two or more circular particles in near or direct contact. The authors applied the model to a modified distributed Lagrange multiplier/fictitious domain method \cite{DT_IJNMF_06}, which allowed the particles to slightly overlap. The model allows the particles to touch each other only at the Kissing stage, but when particles overlap, the virtual steric interaction force kicks in. During the particles overlap, solid surfaces can overlap $1\%$ of the velocity element size. The LF is calculated based on the information on centroid position of different-size particles and their radius, well-constrained $\zeta$, and $\varepsilon$.  $\varepsilon$ is proposed to be chosen based on $\eta$, and the relative density and velocity of the particles with respect to the density and velocity of the fluid, and it is used to limit particles overlap. The LF-3 is suitable for simulating hydrodynamics of different-sized circular particles in near contact or at the collision stage in Newtonian fluids. Numerical simulations with LF-3 agreed with the DKT simulations by Glowinski et al. \cite{GPHJ_JCP_01} closely. Neither of these LF models, however, can be used to simulate close-contact hydrodynamics of non-spherical particles and the LF-3 in \cite{SHJ_IJMFF_03} is not well-suited for simulations with non-Newtonian fluids as it does not account for the effects of the local fluid viscosity on interparticle interactions. 

\begin{figure}[ht!]
\centering
\includegraphics[width=0.95\textwidth]{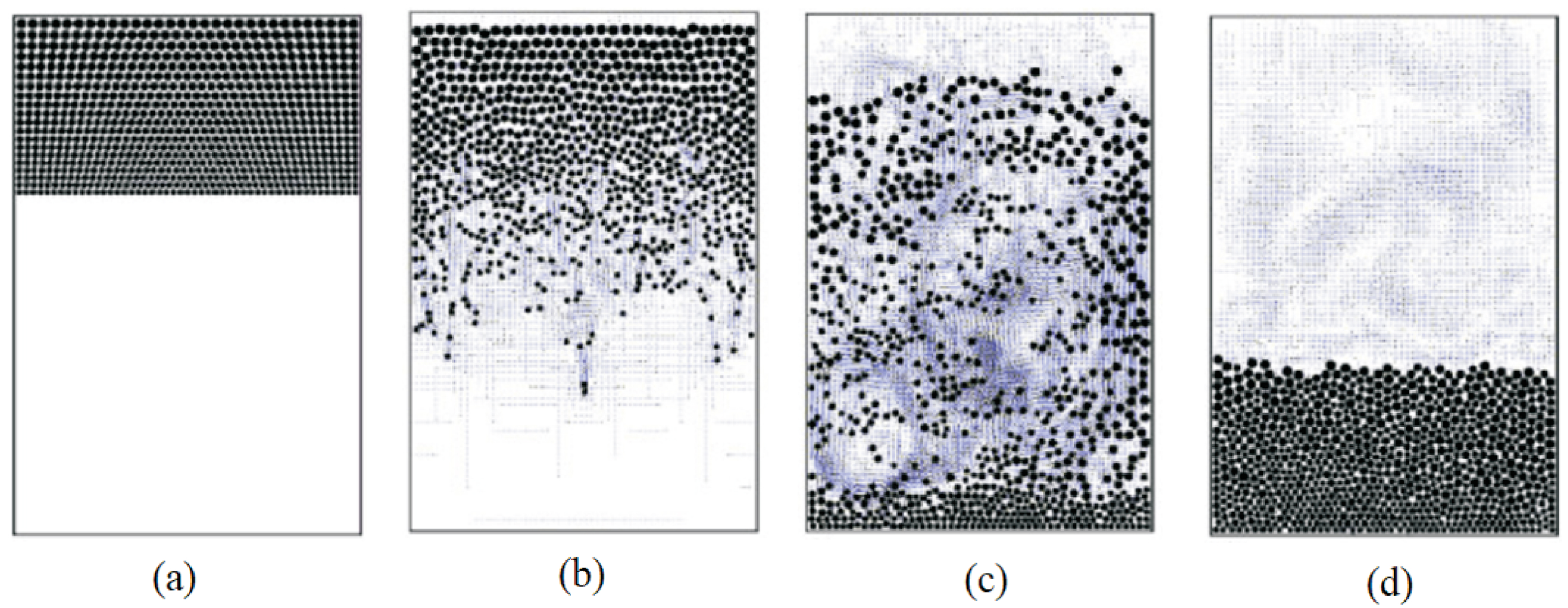}
\caption{Numerical simulation of sedimentation of 790 circular particles of different sizes at non-dimensional times (a) $t$=0, (b) 1.06, (c) 2.68, and (d) 6.19. Interparticle and particle-wall interactions are simulated using Eq. \ref{e20}  (adapted from \cite{DT_IJNMF_06} under conditions of John Wiley and Sons license N. 5355301263951).}
\label{fig:Fig9}
\end{figure}

Wan and Turek \cite{DT_IJNMF_06} introduced an improved lubrication model (LF-4), based on the short-range repulsive model originally developed in \cite{GPHJ_JCP_01, SHJ_IJMFF_03}. The LF-4 allows only small particles to overlap but treats overlapping large particles to be physically inaccurate, therefore, excluding the latter. The model is described as

  \begin{equation}
 \mathbf{F}_{ab}^{pp/w} = \left\{ \begin{array}{rcl}
 0, & |\triangle \mathbf{x}| > R_a + R_b + \zeta , \\
 \frac{1}{\varepsilon} \triangle \mathbf{x} \left( R_a + R_b + \zeta - |\triangle \mathbf{x}| \right)^2, & R_a + R_b <  |\triangle \mathbf{x}| < R_a + R_b + \zeta, \\
  \frac{1}{\varepsilon_p^{\prime}} \triangle \mathbf{x} \left( R_a + R_b - |\triangle \mathbf{x} |\right), & |\triangle \mathbf{x}| \leq R_a + R_b,
\end{array}\right. \label{e20}
 \end{equation}

 \noindent where the second-order accurate $\varepsilon_p^{\prime}$ and the first-order accurate $\varepsilon$ are small positive stiffness parameters. The stiffness parameters for the wall are $0.5\varepsilon'_p$ and $0.5\varepsilon$ \cite{DT_IJNMF_06}. Thus, the LF-4 involves two stiffness parameters to adjust the LF strength at different $\triangle x$. The stiffness parameters are well-constrained if the buoyant force vanishes. A typical range for $\zeta$ is from $0.5\triangle x$ to $2.5\triangle x$. For viscous fluids, $\zeta \sim \triangle x$. If the density ratio between the particle and liquid is on the order of one, then $\varepsilon \sim \left( \triangle x \right)^2 $ and, $\varepsilon_p^{\prime} \sim \triangle x$. The LF-4 is applicable to simulate close-contact dynamics of both circular and non-circular particles with the calculated $\triangle x$. Using Eq. \ref{e20}, Wan and Turek successfully simulated the DKT behavior of two circular particles and settling dynamics of 790 and 3,600 circular particles of various sizes (Fig. \ref{fig:Fig9}). The LF-4, however, does not account for the effects of the local kinematic viscosity on the LF, which is crucial if the fluidic domain is non-Newtonian.
 
 %%%%%%%%%%%%%%%%%%%%%%%%%%%%%%%%%%%%%%%%%%%%%%%%%%%%%%%%%%%%%%%%%%%%%%%%%%%%%%%%%%%%%%%%%%
%----- Figure 9 ----
%%%%%%%%%%%%%%%%%%%%%%%%%%%%%%%%%%%%%%%%%%%%%%%%%%%%%%%%%%%%%%%%%%%%%%%%%%%%%%%%%%%%%%%%%%
%%%%%%%%%%%%%%%%%%%%%%%%%%%%%%%%%%%%%%%%%%%%%%%%%%%%%%%%%%%%%%%%%%%%%%%%%%%%%%%%%%%%%%%%%%
 
Yuan and Ball \cite{YB_JCP_94} introduced a mathematically more complex LF model (LF-5) to simulate  the LF between two equal-sized disk-shaped particles in near contact as a function of $\eta$, which was developed by solving the Stokes equation for the flow domain with periodic boundary conditions. The LF-5 computes normal ($\mathbf{F}_N$) and tangential ($\mathbf{F}_T$) lubrication forces on disk-shaped particles based on decomposed translational velocities in the horizontal and vertical directions and the Rayleigh's dissipation function, representing half of the energy dissipation rate in the fluid flow domain, which is used for viscous force calculations. Because $\mathbf{F}_N$ and $\mathbf{F}_T$ are described as a function of $\mu_f$, the LF-5 is well-suited for simulating near-contact dynamics of particles in non-Newtonian fluids. This model relies on $\zeta$ (not well-constrained), but not on $\varepsilon$. The LF-5 is not suitable to simulate close-contact hydrodynamics of non-spherical particles in bounded domains. Although the LF-5 is mathematically more complex, LF-4 appears to be more flexible to be used for broader applications. 

Ding and Aidun \cite{DA_JSP_03} introduced an improved LF model (LF-6), based on the method described in \cite{ALD_JFM_98}, to enhance the accuracy of hydrodynamic and LF between two solid surfaces in close contact in LBM simulations. In LF calculations, regardless of the presence of a fluid lattice node between solid surfaces in near-contact, a set of virtual nodes is added to the boundary of circular or spherical particles. The model constructs bridge links between the virtual nodes, along which LF is computed based on a force element ($df$) with multiple bridge links given by
 
   \begin{equation}
 df = \left\{ \begin{array}{rcl}
 0, & |\triangle \mathbf{x}| >  \zeta , \\
 \frac{3q}{2c_\sigma^2 \lambda} \nu \rho U \left( \frac{1}{| \triangle \mathbf{x} |^2 } - \frac{1}{c_\sigma^2} \right), & |\triangle \mathbf{x}| \leq  \zeta, 
\end{array}\right. \label{e21}
 \end{equation}

 \noindent where $\rho$ is the particle density and $q$ is the weighing factor, $c_\sigma$ is the scaling factor in the direction of the bridge link, and $u$ is the fluid velocity. For two spherical particles, $\lambda = 0.5(1/R_1+1/R_2)$. The value of q depends on the number and arrangement of the bridge links. Although $q=0.6$ was recommended by the authors, its average value can be computed by $q_{avg} = \sum q df' / \sum df'$ \cite{DA_JSP_03}. The total force and torque on the particle due to the LF are given by 
 
  \begin{equation}
 \mathbf{F}_{lub} = \sum_{BL} df \frac{\mathbf{e}_{\sigma i}}{c_\sigma},
 \label{e22}
 \end{equation}
 
 \begin{equation}
 \mathbf{T}_{lub} = \sum_{BL} \left[ \mathbf{x} - \mathbf{X} \right] \times df \frac{ \mathbf{e}_{\sigma i} }{c_\sigma} ,
 \label{e23}
 \end{equation}

\noindent where $\mathbf{e}_{\sigma i}$ is the velocity vector. This model can be used to simulate close-contact hydrodynamics of particles of different shapes and sizes in Newtonian and non-Newtonian fluids. An enticing feature of LF-6 is that it does not require the presence of a fluid lattice node between solid surfaces in near-contact for LF calculations. The model, however, requires that the grid dimension be smaller than the diameter of the immersed particles. If the bridge link length is smaller than the surface separation distance, the effect of the force element would be negligible. The model involves the threshold distance and the scaling factor for the bridge links. Neither of these parameters is well-constrained.

Although $\varepsilon$ and $\zeta$ are critical parameters in some of the LF models discussed above, their values are typically problem-specific and not well-constrained. To overcome this shortcoming, Diaz-Goano et al. \cite{DGMN_JCP_03} proposed a close-contact interaction model based on repulsive imaginary link without including $\varepsilon$ (LF-7). Different from the other LF models, when the particles come in close contact within $\sim \triangle x$, they move in the direction of the imaginary repulsive link connecting surfaces of neighboring particles at speed proportional to their approaching velocity. As a result, the immersed (Lagrangian) particles change their positions in the intraparticle separation zone in accordance with the correction in the velocity of the immersed particles \cite{DGMN_JCP_03}
 
  \begin{equation}
 \triangle R_{a, b} = \frac{M_{a, b} \left( \zeta - \triangle x_{a,b} \right)}{\left( M_a + M_b\right) },
 \label{e24}
 \end{equation}
 
 \noindent where $M_{a,b}$ is the mass of the particles $a$ or $b$. The velocity correction, $\triangle u_a=\triangle R_a/\delta_t$, is added to the velocity of solid surfaces of the particles in the momentum-balance equation, through which the translation and angular velocities and the new position of the particles are computed at the next time step. Because the LF is dependent on the velocities of approaching particles, it accounts for the effect of the local flow conditions on the LF, as in LF-5. 

\vspace {0.2 cm}

\noindent \textbf{Remarks:} Among the LF models discussed above, the effect of the flow field on the LF calculations is explicitly included in the LF-5 and implicitly accounted for in the LF-7. LF-4 and LF-6 are the only models suitable for simulating close contact and collision dynamics of arbitrary-shaped particles. LF-4 involves two stiffness parameters, which are well-constrained only when the buoyant forces are negligible. LF-6 requires estimates for the threshold distance and the scaling factor for the bridge link, but they are not well-constrained. LF-1, -2, and -7 do not involve the stiffness parameter, but they are limited to simulating close-contact and collision dynamics of circular or spherical particles. LF-1 is applicable for lubrication force calculations between a spherical particle (and possibly uniform-sized multiple spherical particles) and the wall. LF-1, -2, -5, -6, and -7 accommodate the fluid viscosity explicitly in the model formulation; therefore, they are suitable for simulating close-contact dynamics of the particles in non-Newtonian fluids by accommodating the effect of the local kinematic viscosity on the LF. A brief comparison of the LF models is provided in Table \ref{Lubricationforce_Models}.

%%%%%%%%%%%%%%%%%%%%%%%%%%%%%%%%%%%%%%%%%%%%%%%%%%%%%%%%%%%%%%%%%%%%%%%%%%%%%%%%%%%%%%%%%
%---------------------------------------------------------------------
\begin{table}[! htbp]
\centering 
\caption{A comparison of different lubrication force (LF) models.}
\small
\tabcolsep=0.11cm
\begin{threeparttable}
\begin{tabular}{l c c c c c c}
\toprule
\thead{ Model } & \thead{Method} 
 &\thead{Experimental \\ Validation} 
 &\thead{Particles \\ Shape} 
 &\thead{Parameters}
 &\thead{Normal \\ Force}
 &\thead{Tangential \\ Force}
 \\ \midrule

LF-1  &  PIV\tnote{*} \& LBM   & Yes   & Spherical & $\triangle x, \zeta, \eta, U_{\perp}, R_a  $ & Yes & No \\
LF-2  &  \thead{Cluster Implicit, \\  LBM}  & No  & Spherical  & $ \triangle x, \zeta, \eta, R_a, R_b, U_a, U_b $ & Yes  & Yes \\
LF-3  &  \thead{Modified \\ Lagrangian - \\ Eulerian}  & No & \thead{Circular disk}  & $\triangle x, \zeta, \varepsilon, R_a, R_b$ & Yes & Yes \\
LF-4  &  \thead{Fictitious \\ Boundary}  &   No  & \thead{Extendable \\ various shapes} &  $\triangle x, \zeta, \varepsilon_p^{\prime}, \varepsilon, R_a, R_b$ & Yes & No \\
LF-5  &  \thead{Rayleigh's dissipation \\function, modified DLM\tnote{$\dagger$}}   &  Yes  & Spherical &  $\triangle x, \zeta, \varepsilon, R_a$ & Yes & No \\
LF-6  &  \thead{Force element, \\ LBM} &   No  & \thead{Extendable \\ various shapes}  &  \thead{$\triangle x, \zeta, \nu, \rho, U, c_\sigma, \lambda$, \\ $q, R_a, R_b$} & Yes & No \\
LF-7  &  \thead{Fictitious \\ Domain}  &     Yes & Spherical &  $\triangle x, R_a, R_b, M_a, M_b$ & Yes & No \\

\bottomrule\addlinespace[1ex]
\end{tabular}
\begin{tablenotes}\footnotesize
\item[*] Particle Image velocimetry, \item[$\dagger$] Distributed Lagrange Multiplier
\end{tablenotes}
\end{threeparttable}
\label{Lubricationforce_Models}
\end{table}

%%%%%%%%%%%%%%%%%%%%%%%%%%%%%%%%%%%%%%%%%%%%%%%%%%%%%%%%%%%%%%%%%%%%%%%%%%%%%%%%%%%%%
\subsection{Spring Force Models}
\label{sec3.4}
%%%%%%%%%%%%%%%%%%%%%%%%%%%%%%%%%%%%%%%%%%%%%%%%%%%%%%%%%%%%%%%%%%%%%%%%%%%%%%%%%%%%% 

Similar to the LF models, mathematically and physically simple to more complex spring force (SF) models have been developed to simulate short-range hydrodynamic repulsive interactions between particles in near contact. Some of the SF models (e.g., SF–2 discussed below) are similar to the LF models from a mathematical standpoint. But the stiffness parameter in the former is interpreted to be associated with the spring force. In some cases, either a LF model or a SF model can be used to impose the desired short-range repulsive interactions between solid surfaces of the adjacent particles. 
Xu and Michaelides \cite{XM_IJMF_03} proposed a simple SF model (SF-1) to simulate gravity-driven settling and packing of circular particles in a confined domain while allowing the particles to overlap without considering the effect of the interparticle friction. Overlapping particles were simulated using mathematically simple interaction forces with the non-vanishing normal component $F_n^R=\varepsilon \cdot \delta$,  where $\delta$ is the overlapped distance, and the vanishing transverse component, $F_t^R=0$. The time-invariant $\varepsilon$ was proposed to range from 1 to 10, which was set to 5 in numerical simulations. Their simulations showed that interparticle repulsive interactions introduced by SF-1 had an insignificant effect on the simulation results. Another relatively simple SF model (SF-2) was proposed by Pan et al. \cite{PJBG_JFM_02}

  \begin{equation}
 \mathbf{F_i^w} = \left\{ \begin{array}{rcl}
 \frac{C_{ab}}{\varepsilon} \left( \frac{R_a + R_b + \zeta - |\triangle \mathbf{x}| }{\zeta} \right)^2 \frac{\triangle \mathbf{x}}{|\triangle \mathbf{x}|}, & |\triangle x| < R_a + R_b + \zeta , \\
  0, & otherwise.
\end{array}\right. \label{e25}
 \end{equation}

\noindent where $C_{ab}$ is a gravitational force scaling factor, which was set to 1 in this model, therefore the SF-2 is suitable for simulating close-contact dynamics of neutrally-buoyant particles. The threshold interaction distance is well-defined and constrained with respect to the grid spacing. The authors implemented the SF-2 to simulate flow trajectories and velocities of 1,204 spherical particles in a water-filled slit bed using distributed Lagrange multipliers \cite{PJBG_JFM_02}. Similar to the conclusion with the SF-1, the authors concluded based on their combined experimental and numerical analyses that the repulsive force implemented by the SF-2 did not impact overall behavior of the particles. 

Glowinski et al. \cite{GPHJ_JCP_01} proposed a SF model (SF-3) by setting $C_{ab}=\left( \rho_p - \rho_f \right) A_p |\mathbf{g}|$ in Eq. \ref{e25}, where $\mathbf{g}$ is the gravitational acceleration, $A_p$ is the surface area of the particle, subscripts $f$ and $p$ designate the fluid and particle, respectively. Therefore, the SF-3 is well-suited to simulate close-contact dynamics of non-buoyant particles. In simulations with the SF-3 in \cite{GPHJ_JCP_01, GP_Book_19}, $\varepsilon$ and $\zeta$ were set to $\left( \triangle x \right)^2$ and $\triangle x$, in which $\triangle x$ is the lattice spacing. The SF-3 was extended to simulate close-contact dynamics of non-circular and non-spherical particles \cite{HYL_PoF_14, KDDN_JTAC_19} 

  \begin{equation}
 \mathbf{F}_R^{pp} = \left\{ \begin{array}{rcl}
  0, & |\triangle \mathbf{x}_{pp}| > \zeta, \\
 \frac{C_{ab}}{\varepsilon} \left( \frac{1}{\zeta} \right)^2 \left( |\triangle \mathbf{x}_{pp}| - \zeta \right)^2 \frac{\triangle \mathbf{x}_{pp}}{ | \triangle \mathbf{x}_{pp} | }, & |\triangle \mathbf{x}_{pp}| \le \zeta,
\end{array}\right. \label{e26}
 \end{equation}

  \begin{equation}
 \mathbf{F}_R^{pw} = \left\{ \begin{array}{rcl}
  0, & |\triangle \mathbf{x}_{pw}| > \zeta, \\
 \frac{C_{ab}}{\varepsilon} \left( \frac{1}{\zeta} \right)^2 \left( |\triangle \mathbf{x}_{pw}| - \zeta \right)^2 \frac{\triangle \mathbf{x}_{pw}}{ | \triangle \mathbf{x}_{pw} | }, & |\triangle \mathbf{x}_{pw}| \le \zeta.
\end{array}\right. \label{e27}
 \end{equation}
 
\begin{figure}[ht!]
\centering
\includegraphics[width=1.0\textwidth]{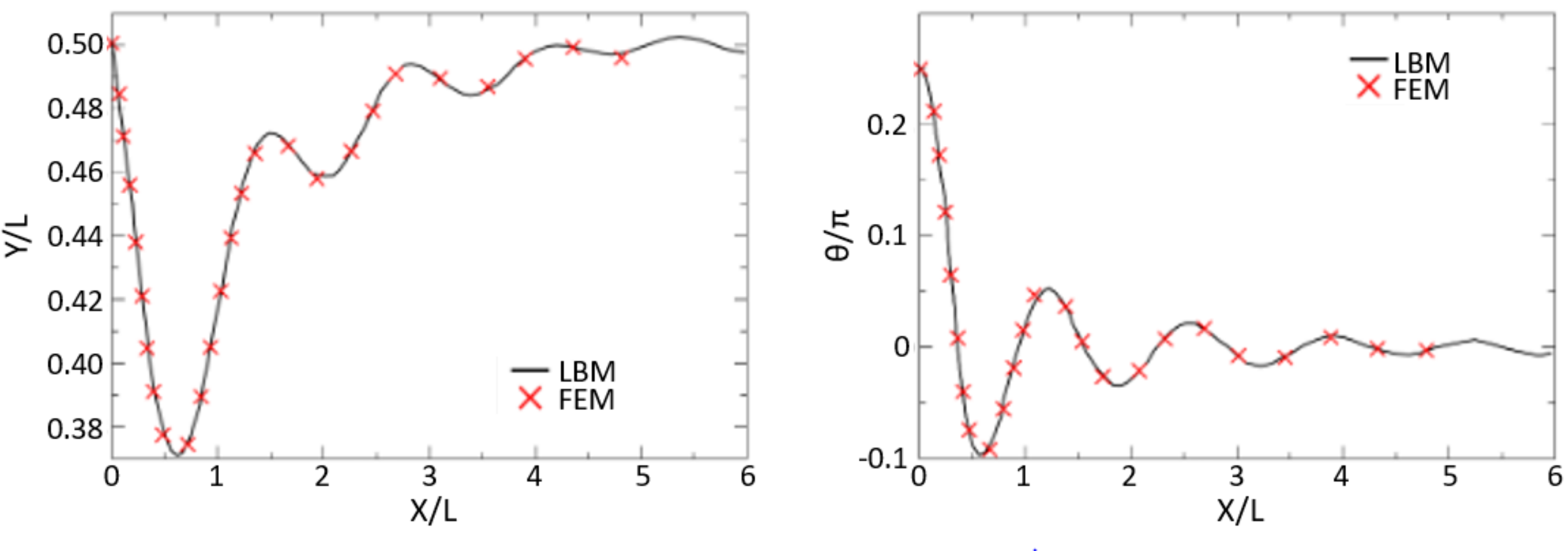}
\caption{Comparison of the particle trajectory and rotations from LBM and FEM simulations using short-range repulsive particle-wall interactions introduced by Eqs. \ref{e26}-\ref{e27} (digitized reconstruction from \cite{XCRY_JFM_09}). }
\label{fig:Fig10}
\end{figure}

 Although the Lagrangian nodes are uniformly distributed on the boundaries of the particles, $\triangle x$ is described differently in the original and extended methods. In the original method, $\triangle x$ is the center-to-center distance between neighboring particles. Similarly, short-range repulsive particle wall interactions are defined based on the distance between the particle center and the closest wall surface. In the extended method, $\triangle x$ is the distance between the Lagrangian nodes on the boundaries of the adjacent particles in close contact or between the Lagrangian nodes on the surface of the particle and the closest wall. In Eqs. \ref{e26} and \ref{e27}, $C_{ab}$ has the dimension of force, if $\varepsilon$ is dimensionless. The SF-3 is suitable to simulate close contact and collision dynamics of arbitrary-shaped particles. Using Eqs. \ref{e26}-\ref{e27}, the settling dynamics of an elliptical particle in a narrow channel filled with the Newtonian fluid were simulated using LBM and FEM. The particle with an initial tilting angle of 45$^\circ$ was released into an initially quiescent fluid from a mid-channel. Fig. \ref{fig:Fig10} show that the particle trajectory and rotations simulated by the LBM or FEM agreed very closely when particle-wall interactions were simulated by Eqs. \ref{e26}-\ref{e27}. 
 
%%%%%%%%%%%%%%%%%%%%%%%%%%%%%%%%%%%%%%%%%%%%%%%%%%%%%%%%%%%%%%%%%%%%%%%%%%%%%%%%%%%%%%%%%%
%----- Figure 10 ----
%%%%%%%%%%%%%%%%%%%%%%%%%%%%%%%%%%%%%%%%%%%%%%%%%%%%%%%%%%%%%%%%%%%

%%%%%%%%%%%%%%%%%%%%%%%%%%%%%%%%%%%%%%%%%%%%%%%%%%%%%%%%%%%%%%%%%%%%%%%%%%%%%%%%%%%%%%%%%%

 This spring force model is applicable only if the particles are denser than the host fluid (i.e., $C_{a,b} > 1$), and hence, is suitable for particle sedimentation problems. The model places a buffer zone around the particles, which prevents their direct contact or collisions. However, it does not accommodate the effect of friction, particle deformability, and lubrication forces, and is limited to simulating close-contact dynamics of the particles without collisions. If the particles cross over the buffer zone, the interparticle repulsive force increases smoothly in proportion to the mass of the particles. The stiffness parameter and threshold interaction distance are well-constrained with respect to the grid spacing. The model does not include the fluid viscosity in the formulation; therefore, it does not address the effect of the local kinematic viscosity on the SF, if the host fluid is non-Newtonian. Some other extensions of this model have been reported in the literature \cite{KDNS_JML_18}.

Feng and Michaelides \cite{FN_JCP_05} formulated a SF model (SF-4) by modifying the short-range repulsive force model of Glowinski et al. \cite{GPHJ_JCP_01, GP_Book_19}. %They used the modified model with the IB-LBM to simulate the gravity-induced settling of particles \cite{JMY_ARB_97}. 
They simulated the settling of 1,232 spherical particles in a narrow channel. In simulations, larger spring forces were applied to restrict particles’ overlapping. The short-range repulsive force between two different-sized spherical particles a and b in near contact is described as \cite{FN_JCP_05} 

 \begin{equation}
\mathbf{F}_{ab}^p = \left\{ \begin{array}{rcl}
 0, & |\triangle\mathbf{x}| > R_a + R_b + \zeta , \\
 \frac{C_{ab}}{\varepsilon} \left( \frac{|\triangle \mathbf{x}| - R_a - R_b - \zeta}{\zeta} \right)^2 \frac{\triangle \mathbf{x}}{|\triangle \mathbf{x}|}, & R_a + R_b <  |\triangle \mathbf{x}| < R_a + R_b + \zeta, \\
  \left[  \frac{C_{ab}}{\varepsilon} \left( \frac{|\triangle \mathbf{x}| - R_a + R_b - \zeta}{\zeta} \right)^2  + \frac{C_{ab}}{\wp_p} \frac{R_a + R_b- |\triangle \mathbf{x}|}{\zeta} \right]   \frac{\triangle \mathbf{x}}{|\triangle \mathbf{x}|}, & |\triangle \mathbf{x}| \leq R_a + R_b,
\end{array}\right. \label{e28}
 \end{equation}
 
\noindent where $\wp_p$ is similar to the stiffness parameter ($\varepsilon$). Eq. \ref{e28} reduces to Eq. \ref{e26} of Glowinski et al. \cite{GPHJ_JCP_01}, if particle overlapping is not considered in simulations. $\wp_p$ in Eq. \ref{e26} should be small enough to strengthen the interparticle or particle-wall repulsive SF to allow small particles to overlap. However, this parameter was not constrained, and hence, left as a calibration parameter. In the sedimentation problem, $\mathbf{F}_{ab}^p$ is smaller prior to packing, but pivotal during the packing stage. Contrary to the results with the SF-1 and SF-2, the short-range interaction forces in the current model were concluded to be influential in multi-body interactions of the particles during their settling and packing.

Kempe and Fröhlich \cite{KF_JFM_12} proposed a physically and mathematically more complex short-range interaction model (SF-5), known as the Adaptive Collision Time Model (ACTM), for IBM. In this model, the total force, ($\mathbf{F}_t$), on the particle has three components, including the sum of the normal collision force ($\mathbf{F}_{n,ab}^{col}$), tangential collision force ($\mathbf{F}_{t,ab}^{col}$), and the lubrication force ($\mathbf{F}_{n,ab}^{lub}$),

  \begin{equation}
\mathbf{F}_t = \sum{\left( \mathbf{F}_{n,ab}^{lub} + \mathbf{F}_{n,ab}^{col} +  \mathbf{F}_{t,ab}^{col}  \right) }.
 \label{e29}
 \end{equation}
 
\noindent The sum of the tangential and normal collision forces is used to correct the pressure and lubrication force on the immersed boundaries. The first component is used to resolve the viscous forces, as the separation distance between the particles becomes less than a lattice unit. The lubrication force for two nonuniform-sized spherical particles $a$ and $b$ in close contact is described by the lubrication model proposed by Brenner \cite{HB_CES_61},
 
 \begin{equation}
\mathbf{F}_{n,ab}^{lub} = \left\{ \begin{array}{rcl}
 0, & 2h \leq |\triangle \mathbf{x}| , \\
 \frac{-6 \pi \nu u_{n,ab}}{|\triangle \mathbf{x}|} \left( \frac{R_a R_b} {R_a +  R_b} \right)^2 \frac{\triangle \mathbf{x}}{|\triangle \mathbf{x}|}, & \zeta_{min} <  |\triangle \mathbf{x}| < 2h, \\
  0, & |\triangle \mathbf{x}| \leq \zeta_{min},
\end{array}\right. \label{e30}
 \end{equation}

\noindent where $\zeta_{min}$ is the minimum separation distance between two particles to avoid singularity due to the coarseness of the immersed boundaries when the particles move into close contact. $u_{n,ab}$ is the relative velocity of the particles.  $\zeta_{min} / \left(2R\right) = 10^{-4}$ and $ h / \zeta_{min} \sim 500$ were recommended. When Eq. \ref{e30} is implemented for a particle near a stationary wall, $R$ in Eq. \ref{e30} is set to be infinite \cite{KF_JCP_12}.  The second component of the total force in Eq. \ref{e29} is the normal collision force. It involves the spring force, based on the theory Hertz with a dissimilar coefficient ($k_n$), and material damping with the coefficient of damping ($\eta_n$) \cite{KF_JCP_12, FN_JCP_05} 
 
  \begin{equation}
 \mathbf{F}_{n,ab}^{col} = \left\{ \begin{array}{rcl}
  0 , & 0 < |\triangle \mathbf{x}| , \\
  -\left( k_n |\triangle \mathbf{x}|^{3/2} + \eta_n u_{n,ab}\right) \mathbf{n}_{ab}, & |\triangle \mathbf{x}| \leq 0.
\end{array}\right. \label{e31}
 \end{equation}

\noindent Here, $\eta_n$ and $k_n$ are calculated iteratively to match the experimentally or theoretically-determined reference rebound velocity and the collision time. The collision time was an order of magnitude larger than the maximum time-step in their simulations. Using Eq. \ref{e30}, the IBM results closely matched experimentally obtained trajectories and velocities of the particles in \cite{GLP_PoF_02}. $\mathbf{F}_{t, ab}^{col}$ requires two experimentally determined parameters; the effective impact angle ($\Psi_{in}^{RS}$) associated with the flow mode of the particles (rolling and sliding) and the friction coefficient ($\sigma_f$) associated with the sliding motion of the particles. The local impact angle is critical for the sliding motion of the particles and is defined as $\Psi_{in} = \mathbf{U}_{t,ab}^c / ( \mathbf{u}_{t,ab} \cdot \mathbf{n}_{ab}) $, in which  $\mathbf{u}_{t,ab}$ is the relative velocity of the center of the particles in the tangential direction $t$ ($\mathbf{u}_{t,ab}=\mathbf{u}_b-\mathbf{u}_a$) and $\mathbf{U}_{t,ab}^c$ is the particle velocity at the point of contact. $\mathbf{U}_{t,ab}^c$ and $\mathbf{u}_{t,ab}$ are computed as \cite{KF_JCP_12, FN_JCP_05} 

  \begin{equation}
\mathbf{u}_{t,ab} = \mathbf{u}_{ab} - \left( \mathbf{n}_{ab} \cdot  \mathbf{u}_{ab} \right) \mathbf{n}_{ab},
 \label{e32}
 \end{equation}

  \begin{equation}
\mathbf{U}_{t,ab}^c = \mathbf{u}_{ab} + R_a \left( \mathbf{\omega}_{a} \times \mathbf{n}_{a,b}\right)  - R_b \left( \mathbf{\omega}_{b} \times \mathbf{n}_{ab}\right).
 \label{e33}
 \end{equation}
 
 \noindent When the local impact angle is greater than the effective impact angle, the particle slides. In this case, the resultant tangential force is computed by \cite{FN_JCP_05} 
 
  \begin{equation}
\mathbf{F}_{t,ab}^{col} = \sigma_f |\mathbf{F}_n|\mathbf{t}_{ab}^c.
 \label{e34}
 \end{equation}
 
 \noindent Conversely, if the local impact angle becomes smaller than the effective impact angle, the relative surface velocity at the contact surface has to be zero and the particle displays a rolling motion.

\vspace {0.2 cm}

\noindent \textbf{Remarks:} The SF-1 is a simple SF model, in which SF relies on the interaction strength introduced by the stiffness parameter and the overlapped area of the circular particles. However, neither of them is well-constrained. The SF-3 is applicable to sedimentation problems.. The model places a buffer zone around the particles, which prevents their direct contact or collisions. However, it does not accommodate the effect of friction, particle deformability, and lubrication forces, and is limited to simulating close-contact dynamics of the particles without collisions. In comparison to SF-3, SF-2 can be used to simulate close-contact hydrodynamics of neutrally-buoyant, nonuniform-size spherical particles, and hence, is suited for simulating hydrodynamic interactions of the particles in a flowing fluid. Unlike the SF-1, but similar to the SF-3, the stiffness and the threshold interaction distance in SF-2 are well-defined and constrained with respect to the grid spacing. SF-4 reduces to SF-3 (Eq. \ref{e28} vs. Eq. \ref{e26}) if particle-overlapping is not considered in simulations. $\wp_p$ in SF-4 should be small enough to strengthen the interparticle or particle-wall repulsive spring forces to allow small particles to overlap. However, this parameter was not constrained, and hence, left as a calibration parameter. In the sedimentation problem, $\mathbf{F}_{ab}^p$ is smaller prior to packing, but pivotal during the packing stage. Contrary to the SF-1 and SF-2, the short-range interaction forces in the current model were concluded to be influential in multi-body interactions of the particles during their settling and packing. Different from previously discussed interaction models, the SF-5 involves both the lubrication and collision forces, in which the lubrication force component is built on the SF model and material damping. Unlike the previously discussed models, it accommodates the effect of the rolling or sliding motion of the particle in the collision force calculations. The collision forces, however, require problem-specific value for the effective impact angle friction that needs to be determined experimentally. The interaction distances for the lubrication force and collision force calculations are constrained. The model in its current version is limited to short-range repulsive interactions of spherical particles. Among all these SF models, SF-5 includes the effect of the fluid viscosity in interparticle interaction force calculations, and hence, is suitable for simulating particles in non-Newtonian fluids. A concise comparison of the spring-based interaction models is provided in Table \ref{Springforce_Models}.      

 %%%%%%%%%%%%%%%%%%%%%%%%%%%%%%%%%%%%%%%%%%%%%%%%%%%%%%%%%%%%%%%%%%%%%%%%%%%%%%%%%%%%%%%%%%
%---------------------------------------------------------------------
\begin{table}[! htbp]
\centering 
\caption{A comparison of different spring force (SF) models.}
\small
\tabcolsep=0.11cm
\begin{threeparttable}
\begin{tabular}{l c c c c c c}
\toprule
\thead{ Model } & \thead{Method} 
 &\thead{Experimental \\ Validation} 
 &\thead{Particles \\ Shape} 
 &\thead{Parameters}
 &\thead{Normal \\ Force}
 &\thead{Tangential \\ Force}
 \\ \midrule

SF-1  &  LBM   & No   & Circular & $\varepsilon, \triangle x, \delta$ & Yes & No \\
SF-2  &  DLM   & Yes  & Spherical  & $\varepsilon, \triangle x, \zeta, C_{ab}, R_a, R_b$ & Yes  & No \\
SF-3  &  \thead{Fictitious \\ domain/DLM}  & Yes & \thead{Extendable to \\various shapes}  & $\varepsilon, \triangle x, \zeta, C_{ab}, R_a, R_b $ & Yes & No \\
SF-4  &  IB-LBM  &   Yes  & Spherical &  $\varepsilon, \triangle x, \zeta, C_{ab}, \wp_p, R_a, R_b$,  & Yes & No \\
SF-5  &  IBM   &     Yes & Spherical &  $\triangle x, \zeta_{min}, \mu_f, \sigma_f, \eta, R_a, R_b$ & Yes & No \\

\bottomrule\addlinespace[1ex]
\end{tabular}
\end{threeparttable}
\label{Springforce_Models}
\end{table}
%---------------------------------------------------------------------

%%%%%%%%%%%%%%%%%%%%%%%%%%%%%%%%%%%%%%%%%%%%%%%%%%%%%%%%%%%%%%%%%%%%%%%%%%%%%%%%%%%%%%%%%%

%%%%%%%%%%%%%%%%%%%%%%%%%%%%%%%%%%%%%%%%%%%%%%%%%%%%%%%%%%%%%%%%%%%%%%%%%%%%%%%%%%%%%
\subsection{Hard-sphere and Soft-sphere Models}
\label{sec3.5}
%%%%%%%%%%%%%%%%%%%%%%%%%%%%%%%%%%%%%%%%%%%%%%%%%%%%%%%%%%%%%%%%%%%%%%%%%%%%%%%%%%%%% 
 Hard-sphere and soft-sphere models can be used to simulate close-contact and collision dynamics of deformable or rigid particles in shear flow considering the friction force among them. These two models are commonly used with the DEM \cite{MS_IJMF_92, MS_IJMF_01}, and have also been implemented with the LBM and IBM to simulate close-contacts and collision dynamics of the particles in granular systems \cite{ZTSN_CF_14, HFO_CS_07, FHO_IJNME_07}.

 %%%%%%%%%%%%%%%%%%%%%%%%%%%%%%%%%%%%%%%%%%%%%%%%%%%%%%%%%%%%%%%%%%%%%%%%%%%%%%%%%%%%%
\subsubsection{Hard-sphere Models}
\label{sec3.5.1}
%%%%%%%%%%%%%%%%%%%%%%%%%%%%%%%%%%%%%%%%%%%%%%%%%%%%%%%%%%%%%%%%%%%%%%%%%%%%%%%%%%%%% 
 
The hard-sphere model (HSM) was coupled with the DPM for the first time by Campbell and Brennen \cite{CB_JFM_85} to simulate the distribution of density, velocity, and temperature in a 2D Couette flow in response to unidirectional flow of inelastic particles. The model is computationally efficient, if the number of particles is small, but becomes less efficient if the number of particles increases \cite{CB_JFM_85}. The HSM could also be computationally-inefficient if inelastic collapses occur in simulations \cite{HB_CES_61}, in which the kinetic energy arises from the collision of particles dissipates in the fluid. Using the HSM, trajectories of the particles are calculated from momentum-conserving pair-wise additive binary collisions \cite{HKB_CES_96}. Wang et al. \cite{WZWX_Parti_10} used the hybrid HSM-LBM to simulate 2D particulate flow in a gas-filled domain, in which IBM was used to couple the solid and gas phases. The authors simulated the sedimentation of 1,400 particles, which were 2.5 or 1,500 denser than the gas in a fluidized bed. The standard HSM relies on the impulse momentum ($J$) that accounts for the mutual repulsion of colliding surfaces of two spherical particles \cite{NZSM_Book_16, RD_Book_11, OASN_Conf_19}. The normal component of $\mathbf{J}$ ($J_n$) is described as  
 
  \begin{equation}
J_n = -m_{eff} \left( 1+ e_n \right) \left( \mathbf{n}_{ab} \cdot \mathbf{V}_{ab}^{0} \right),
 \label{e35}
 \end{equation}
 
   \begin{equation}
m_{eff} = \frac{m_a m_b}{m_a+m_b},
 \label{e36}
 \end{equation}
 
\noindent where $\mathbf{V}_{ab}^{0}$ is the pre‐collision relative velocity and $e_n$ is the normal coefficient of restitution that represents absolute momentum loss during intraparticle inelastic collision. The superscript 0 represents the pre-collision quantities. The ratio of post- to pre-collision velocity varies between 0 and 1 in the absence of external forces. The value of $e_n$ is empirically determined and typically assumed to be constant. The tangential component of $\mathbf{J}$ ($J_t$) is associated with sticking collision when $\mathbf{V}_n > \mathbf{V}_t$, as described by Eqs. \ref{e35} and \ref{e36}, respectively,

   \begin{equation}
J_t = -\nu J_n  \quad for \quad \nu J_n < \frac{2}{7} \left(1+e_t \right) m_{eff} \left(\mathbf{t}_{ab} \cdot \mathbf{V}_{ab}^0 \right),
 \label{e37}
 \end{equation}

    \begin{equation}
J_t = -\frac{2}{7} \left(1+e_t \right) m_{eff} \left(\mathbf{t}_{ab} \cdot \mathbf{V}_{ab}^0 \right)  \quad for \quad \nu J_n \geq \frac{2}{7} \left(1+e_t \right) m_{eff} \left(\mathbf{t}_{ab} \cdot \mathbf{V}_{ab}^0 \right),
 \label{e38}
 \end{equation}

\noindent where $e_t$ is the tangential coefficient of restitution. After $J$ is computed, post-collision translational and angular velocities of the particles - denoted by superscript 1 - are calculated by 
 
     \begin{equation}
m_a \mathbf{V}_{a}^1 = m_a \mathbf{V}_{b}^0 + \mathbf{J}; \quad m_b \mathbf{V}_{b}^1 = m_b \mathbf{V}_{b}^0 - \mathbf{J},
 \label{e39}
 \end{equation}
 
      \begin{equation}
I_a \mathbf{\omega}_{a}^1 = I_a \mathbf{\omega}_{a}^0 + R_a\mathbf{n}_{ab} \times \mathbf{J}; \quad I_b \omega_{b}^1 = I_b \omega_{b}^0 + R_a\mathbf{n}_{ab} \times \mathbf{J},
 \label{e40}
 \end{equation}
 
       \begin{equation}
\mathbf{J} = J_n \mathbf{n}_{ab} + J_t \mathbf{t}_{ab},
 \label{e41}
 \end{equation}
 
 \noindent where $I_a$ is the moment of intertia. This HSM is derived for spherical particles in the absence of external forces. The normal and tangential coefficients of restitution depend on the properties of the particle and contact conditions. The resultant momentum transfers are described by normal and tangential components, whose relative contributions vary with whether the collision is in the form of sticking or sliding. Post-collision velocities of the particles are computed from elastic intraparticle collisions. 
 
 %%%%%%%%%%%%%%%%%%%%%%%%%%%%%%%%%%%%%%%%%%%%%%%%%%%%%%%%%%%%%%%%%%%%%%%%%%%%%%%%%%%%%
\subsubsection{Soft-sphere Models }
\label{sec3.5.2}
%%%%%%%%%%%%%%%%%%%%%%%%%%%%%%%%%%%%%%%%%%%%%%%%%%%%%%%%%%%%%%%%%%%%%%%%%%%%%%%%%%%%% 
 
 Soft-sphere models (SSMs) allow simulations to run at a fixed time-step and have been used with DPMs. Cundall and Strack \cite{CS_Geotech_79} developed the SSM that allows the deformable or rigid particles to overlap. The overlapped distance with respect to the velocity of the particles is used to calculate interparticle contact forces. The SSMs are flexible enough to accommodate other short-range interaction forces such as van der Waals or electrostatic forces. 
 
 %%%%%%%%%%%%%%%%%%%%%%%%%%%%%%%%%%%%%%%%%%
 \vspace{0.2 cm}
 \noindent \textbf{Soft-sphere Model - 1 }
 
 Breugem \cite{WPB_Conf_10} combined the SSM with the second-order IBM, developed by Uhlmann \cite{MU_JPC_05}, to simulate the collisions of spherical particles. Using the split-direct forcing approach suggested by Lue et al. \cite{LWFC_PRE_07}, Breugem \cite{WPB_Conf_10} improved the accuracy of the forcing technique in the IBM. Unlike the LF-3, tangential forces were not considered in the collision process. The model calculates the normal contact forces between the particles $a$ and $b$ by \cite{WPB_Conf_10} 
 
\begin{equation}
\mathbf{F}_{ab,n} = -k_n \delta_n \mathbf{n}_{ab} - \eta_n \mathbf{U}_{ab,n}; \quad k_n = \frac{m_e \left( \pi^2 + \left[ \ln e_d\right]^2 \right)}{\left[ \triangle t N_c\right]^2}; \quad \eta_n = -\frac{2m_e \left[ \ln e_d\right]^2}{\triangle t N_c},
 \label{e42}
 \end{equation}
 
 \begin{equation}
m_e = \frac{1}{\frac{1}{m_a} + \frac{1}{m_b} },
 \label{e43}
 \end{equation}
 
 \noindent where $m_e$ is effective mass, $\triangle t N_c$ is contact time, $k_n$ is the normal spring stiffness, and $\eta_n$ is the normal damping coefficient. $e_d$ is the restitution coefficient defined as $-d\delta_n/dt|_{t=\triangle t N_c}/-d\delta_n/dt|_{t=0}$  for dry collision. $\mathbf{F}_{ab,n}$ is defined as a function of the position and velocity of the particle at the next time step ($n+1$), which can be obtained iteratively through the under-relaxation scheme by setting the relaxation factor to 0.5 \cite{WPB_Conf_10}. Using the SSM, close contact and collision dynamics of two equal-sized spherical particles with the radius of $R$ near a smooth wall were simulated in \cite{WPB_Conf_10}. Three scenarios were considered, involving i) two particles were initially positioned away from the wall with an initial separation distance of 0.25$R$, ii) two particles were initially positioned side by side and away from the wall with an initial separation distance of 0.25$R$, and iii) a particle approaching a wall. Simulations with the coarse mesh became unstable as the distance between solid surfaces ($\triangle x$) became less than one lattice unit. In the second scenario, two simulations involving a non-rotating spherical particle or a freely rotating spherical particle were performed. The authors noted that no lubrication correction was needed for the freely rotating spherical particle. However, the lubrication model for the non-rotating spherical case is given by \cite{WPB_Conf_10}
 
 \begin{equation}
\triangle \mathbf{F}_{pp} = -6\pi\mu_f R \mathbf{u}_c \left[ \lambda_{pp} \left( \Psi \right) \right) - \lambda_{pp} \left( \Psi_{pp} \right)],
 \label{e44}
 \end{equation}
 
  \begin{equation}
 \lambda_{pp} \left( \Psi \right) = \frac{1}{2\Psi}-\frac{9}{20} \log \Psi - \frac{3}{56}\Psi \log \Psi + 1.346 + O \left( \Psi \right),
 \label{e45}
 \end{equation}
 
 \noindent where $\lambda$ is the Stokes amplification factor, and $\Psi_{pp}$ =0.025. Similarly, the lubrication force for the third scenario was needed when the particle was within the threshold interaction distance away from the wall, which was computed by \cite{WPB_Conf_10}
 
  \begin{equation}
\triangle \mathbf{F}_{pw} = -6\pi\mu_f R \mathbf{U}_c \left[ \lambda_{pw} \left( \Psi \right) \right) - \lambda_{pw} \left( \Psi_{pw} \right)],
 \label{e46}
 \end{equation}
 
   \begin{equation}
 \lambda_{pp} \left( \Psi \right) = \frac{1}{\Psi}-\frac{1}{5} \log \Psi - \frac{1}{21}\Psi \log \Psi + 0.9713 + O \left( \Psi \right),
 \label{e47}
 \end{equation}
 
\noindent where $\mathbf{U}_c$ is the particle velocity, $\Psi_{pw}$=0.075 and $\Psi=\triangle x / R$. Additional forces and torques associated with the interparticle and particle-wall interaction forces computed by the SSM are added to the total forces and torques on the particles to update their translation and angular velocities. 
This soft-sphere model allows the particles to collide and overlap. When it is used in conjunction with the lubrication forces, a lubrication correction may be required for non-rotating particles. The model is applicable to rigid or deformable particles.

\vspace{0.2 cm}
 %%%%%%%%%%%%%%%%%%%%%%%%%%%%%%%%%%%%%%%%%%
 \noindent \textbf{Soft-sphere Model - 2 }
 
 Feng et al. \cite{FMM_JFE_10} studied collisions of spherical particles with a wall using the DPM and the DNS (IBM) in a viscous fluid. They used a SSM-based collision model proposed by Cundall and Strack \cite{CS_Geotech_79} to handle collisions between solid surfaces, as the HSMs (e.g., the HSM by Alder and Wainwright \cite{AW_JCP_57}) do not allow solid surfaces to overlap. The slip and rebound velocities near solid surfaces are determined by the collision parameters (e.g., spring stiffness and damping coefficient). Oblique collisions that occur when the velocity vector of any two solid surfaces has a non-zero angle with respect to the collision line are also handled by the model. The linear dashpot method was used to derive the tangential ($t$) and normal forces ($n$). The normal and tangential collision forces between solid surfaces of the particles $a$ and $b$ are computed by \cite{FMM_JFE_10}

    \begin{equation}
 f_{ab}^n=-\varepsilon_n \delta_{ab}^n - \Im_n u_{ab}^n ; \quad
 f_{ab}^t=-\varepsilon_t \delta_{ab}^t - \Im_t u_{ab}^t.
 \label{e48}
 \end{equation}

\noindent  where $u_{ab}^n$ and $u_{ab}^t$ are the normal and tangential components of the relative velocity of the particles $a$ and $b$, $\varepsilon$ and $\Im$ are the normal spring stiffness and damping coefficients,  $t$ and $n$ correspond to their normal and tangential components, and $\delta_{ab}^n$ and $\delta_{ab}^t$ are the overlapping displacements in the normal and tangential directions. The relative tangential velocity ($\mathbf{u}_{ab}^t$) can be computed as \cite{AW_JCP_57}
 
     \begin{equation}
\mathbf{u}_{ab}^t = \mathbf{u}_{ab} - \left( \mathbf{u}_{ab}  \cdot \mathbf{n}_{ab}  \right) \mathbf{n}_{ab} - \left[ \omega_a \times R_a \mathbf{n}_{ab} - \omega_b \times R_b\left( -  \mathbf{n}_{ab} \right)  \right]
 \label{e49}
 \end{equation}

  \begin{equation}
 f_{ab}^{t} = \left\{ \begin{array}{rcl}
  -k_t \delta_{ab}^t - \eta_t u_{ab}^t, & |f_{ab}^{t}| \leq \mu_s |f_{ab}^{n}|, \\
 \mu_k |f_{ab}^{n}|\frac{\delta_{ab}^t}{\delta_{ab}^t}, & \mu_s  |f_{ab}^{n}| < |f_{ab}^{t}|,
\end{array}\right. \label{e50}
 \end{equation} 
 
\noindent where $\mu$ is the coefficient of friction, and subscripts $s$ and $d$ refer to the mean static (associated with immobile particles) and dynamic friction (associated with mobile particles), respectively. The same value of $\mu$ is typically chosen for static friction and dynamic friction, and it was set to 0.3 in simulations in \cite{FMM_JFE_10}. Feng et al. \cite{FMM_JFE_10} validated their numerical model, in which particle-wall interactions were simulated using the SSM against the experimentally determined approaching and rebounding velocities of a particle during the central-wall collision by Jospeh et al. \cite{JZR_JFM_01}. The authors demonstrated that the simulated results closely matched the experimental data, validating the accuracy of the SSM in particle-wall collision simulations. This model is capable of simulating oblique collisions. Stiffness parameter affects the collision duration, and soft springs may facilitate considerable overlapping between the surfaces of the particles. Larger $k_t$ results in smaller time-steps. Typically, $200 \leq k_t < 5\times 10^4$ (N/m) and $0 \leq \eta < 0.1$ (Ns/m).

 SSM and HSM have been commonly used to simulate particulate flows especially in granular systems. SSM can handle multiple interparticle contacts and overlapped surfaces. Collisions and interactions between bodies are deemed to be binary and instantaneous in HSM \cite{NZSM_Book_16,DAVK_CES_07}. If the number of particles is small, HSM is computationally more efficient and faster than SSM, as the contact time in the SSM would be considerably longer than the time-step in simulations.

\vspace {0.2 cm}

\noindent \textbf{Remarks:} SSM and HSM have been commonly used to simulate particulate flows especially in granular systems. Each model offers distinct advantages in different problem set-ups. If the number of particles is small, the HSM is computationally more efficient and faster than SSM, as the contact time in the SSM would be considerably longer than the time-step in simulations. On the other hand, the SSM is capable of simulating multiple interparticle contacts and overlapped surfaces. In addition, in the HSM formulation, collision and interaction between bodies are deemed to be binary and instantaneous \cite{NZSM_Book_16,DAVK_CES_07}. Table \ref{Sphere_Models} summarizes the main features of each modeling approach.   

  %%%%%%%%%%%%%%%%%%%%%%%%%%%%%%%%%%%%%%%%%%%%%%%%%%%%%%%%%%%%%%%%%%%%%%%%%%%%%%%%%%%%%%%%%%
%---------------------------------------------------------------------
\begin{table}[! htbp]
\centering 
\caption{A comparison of the performance of a soft-sphere model against a hard-sphere model.}
\small
\tabcolsep=0.11cm
\begin{threeparttable}
\begin{tabular}{l c c }
\toprule
\thead{ Features } & \thead{Soft-sphere Model} 
 &\thead{Hard-sphere Model} \\ \midrule

Multiple interaction       &  No   & No     \\
Non-uniform particle size  &  Yes  & No     \\
Computational productivity &  Yes  & Yes    \\
Dense regime               &  Yes  & No     \\
Dilute regime 	           & Yes   & Yes    \\
Time step sensitivity	   & Yes   & Yes    \\
Deformation	               & Yes   & No     \\
 
\bottomrule\addlinespace[1ex]
\end{tabular}
%\begin{tablenotes}\footnotesize
%\item[*] Missing fraction of the particular data type in the available climate records 
%\end{tablenotes}
\end{threeparttable}
\label{Sphere_Models}
\end{table}
%---------------------------------------------------------------------
 
%%%%%%%%%%%%%%%%%%%%%%%%%%%%%%%%%%%%%%%%%%%%%%%%%%%%%%%%%%%%%%%%%%%%%%%%%%%%%%%%%%%%%
\subsection{Lennard-Jones Potentials}
\label{sec3.6}
%%%%%%%%%%%%%%%%%%%%%%%%%%%%%%%%%%%%%%%%%%%%%%%%%%%%%%%%%%%%%%%%%%%%%%%%%%%%%%%%%%%%% 
 
 The Lennard-Jones (LJ) potentials have been implemented to simulate steric and/or attractive interaction forces between particles and particles and stationary solid objects \cite{DNKK_PhyA_16, YP_JCIS_04}. The LJ potential energy can be written as a function of the center-to-center separation distance, $|\triangle \mathbf{x}|$, if the pair-wise interaction between two particles is independent of their interactions with other particles. As an example, the LJ 6-12 potential energy, $E_{ab}$, between two particles $a$ and $b$ is given by \cite{YP_JCIS_04}

\begin{equation}
 E_{ab}=0.4\varepsilon \left[ \left( \frac{R_a+R_b}{|\triangle \mathbf{x}|}\right)^{12} - \left(  \frac{R_a+R_b}{|\triangle \mathbf{x}|} \right)^6    \right],
 \label{e51}
 \end{equation}
 
\noindent where $\varepsilon$ is the potential depth. The LJ 6-12 forces for the particle-particle and particle-wall interactions can be obtained from Eq. \ref{e51} \cite{DNKK_PhyA_16} 
 
  \begin{equation}
 \mathbf{F}^{pp} = \left\{ \begin{array}{rcl}
  0, & |\triangle \mathbf{x}| > R_a + R_b + \zeta, \\
2.4 \varepsilon \sum_{b=1, a \neq b}^{N} \left[2\left( \frac{R_a + R_b}{|\triangle \mathbf{x}|} \right)^{14} - \left( \frac{R_a + R_b}{|\triangle \mathbf{x}|} \right)^8 \right] \frac{\mathbf{x}_c^a - \mathbf{x}_c^b}{\left( R_a + R_b\right)^2}, & |\triangle \mathbf{x}| \leq R_a + R_b+ \zeta,
\end{array}\right. \label{e52}
 \end{equation} 
 
\begin{equation}
 \mathbf{F}^{pw} = \left\{ \begin{array}{rcl}
  0, & |\triangle \mathbf{x}| > 2R_a  + \zeta, \\
2.4 \varepsilon \sum_{j=1}^{jmax} \left[2\left( \frac{R_a}{|\triangle \mathbf{x}|} \right)^{14} - \left( \frac{R_a}{|\triangle \mathbf{x}|} \right)^8 \right] \frac{\mathbf{x}_c^a - \mathbf{x}_w^j}{\left( R_a \right)^2}, & |\triangle \mathbf{x}| \leq 2R_a + \zeta,
\end{array}\right. \label{e53}
 \end{equation} 
 
\begin{figure}[ht!]
\centering
\includegraphics[width=1.\textwidth]{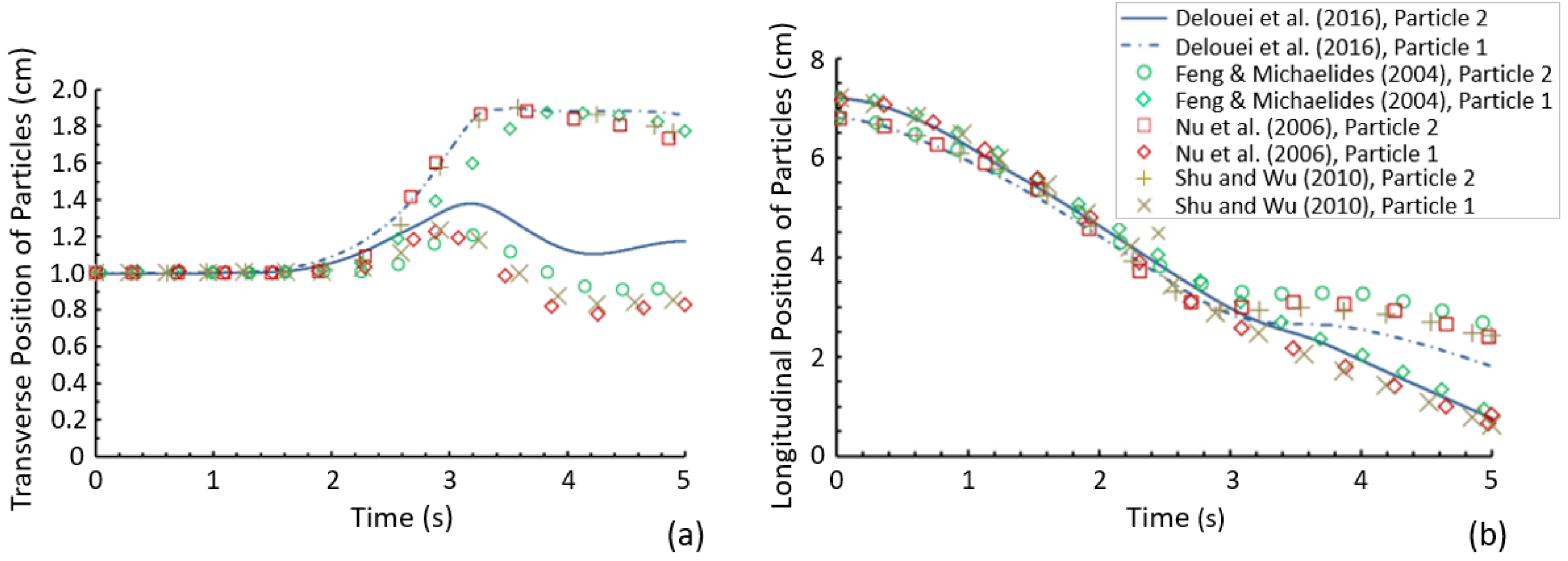}
\caption{Time histories of the transverse (perpendicular to the settling direction) position and (b) the longitudinal (in the settling direction) position of two circular particles settling in a Newtonian fluid under gravity, in comparison to findings from the earlier studies (from Ref. \cite{DNKK_PhyA_16} under conditions of Elsevier license N. 5355280607026).}
\label{fig:Fig12}
\end{figure}

\noindent This LJ model does not involve calibration parameters \cite{HLSN_PRE_15}. The positive and negative interaction forces correspond to the attractive and repulsive interactions, respectively. The recommended values for the stiffness parameter ($\varepsilon$) and the threshold interaction distance ($\zeta$) are reported to be $R^2$ and one lattice unit, respectively \cite{DNK_CiCP_16}. The model is commonly used to simulate the collisions of large particles (a few microns in size) in solvents \cite{BKP_Book_06}. The LJ 6-12 forces were used by Amiri Delouei et al. \cite{DNKK_PhyA_16} to implement interparticle and particle-wall interactions in simulating gravity-driven settling of two circular particles in a bounded domain filled with Newtonian fluid. Fig. \ref{fig:Fig12} shows that the transverse and longitudinal positions of the particles simulated by Amiri Delouei et al. \cite{DNKK_PhyA_16} are in good agreement with the findings from the earlier studies (Fig. \ref{fig:Fig12}). The difference between the results after $\sim2.8$ sec can be attributed to the unstable configuration of the particles in the flow direction at the tumbling stage.

%%%%%%%%%%%%%%%%%%%%%%%%%%%%%%%%%%%%%%%%%%%%%%%%%%%%%%%%%%%%%%%%%%%%%%%%%%%%%%%%%%%%%%%%%%
%----- Figure 11 ----
%%%%%%%%%%%%%%%%%%%%%%%%%%%%%%%%%%%%%%%%%%%%%%%%%%%%%%%%%%%%%%%%%%%

%%%%%%%%%%%%%%%%%%%%%%%%%%%%%%%%%%%%%%%%%%%%%%%%%%%%%%%%%%%%%%%%%%%%%%%%%%%%%%%%%%%%%%%%%%

Ba\c sa\u gao\u glu and Succi \cite{BS_JCP_10} implemented LJ 6-12 interaction potentials in a colloidal LBM model to simulate steric interactions between the boundary nodes of neighboring circular particles (i.e., they implemented surface-to-surface rather than center-to-center repulsive interactions) and between the boundary nodes of the particles and walls. They used the LJ model to simulate steric interaction-induced temporary and permanent entrapments of particles in a rough-walled narrow flow channel. They used the simplified version of the two-body Lennard-jones force ($F_{L,J}$) \cite{FY_PRE_08} to simulate repulsive interactions between solid surfaces of the objects in close proximity \cite{BS_JCP_10}.

\begin{equation}
 \mathbf{F}_{LJ}=\varepsilon \left( \frac{|\triangle \mathbf{x} | }{|\triangle \mathbf{x}|_i} \right)^{-13} \mathbf{n},
 \label{e54}
 \end{equation}

 \noindent where $|\triangle \mathbf{x}|$ is the separation distance between boundary nodes on the solid surfaces of two objects (particle-particle or particle-wall), $|\triangle \mathbf{x}|_i$ denotes the threshold interaction distance beyond which steric interactions vanish, and $\mathbf{n}$ is the unit vector in the direction of steric interaction forces. By changing the stiffness parameter ($\varepsilon$), the authors simulated three distinct particle behaviors of two-particle flow simulations in 2D, involving (i) temporary entrapment of two-particles, (ii) temporary entrapment of one of the particles and permanent entrapment of the other particle, and (iii) permanent entrapment of both particles. Their simulations revealed that any irregularity on wall surfaces could cause temporary or permanent clogging of the flow field if hydrodynamic forces on the particle cannot overcome unevenly distributed repulsive barrier on the channel wall. Ba\c sa\u gao\u glu et al. used the LJ 6-12 model to simulate chemotatic motility \cite{DBNH_Entropy_19} or flow \cite{BHNS_CPC_17, BHNS_CPC_17_2} of circular particles in a Newtonian fluid, settling or flow of a mixture of non-circular particles in a Newtonian fluid, and flow of a mixture of non-circular particles in a non-Newtonian fluid \cite{BBSF_MN_19, BSWB_SR_18}. Using the LJ 6-12 potentials, Ba\c sa\u gao\u glu et al. \cite{BSWB_SR_18} demonstrated that steady vortices do not necessarily always control particle entrapments nor do larger particles get selectively entrapped in steady vortices in a microfluidic chamber. Interestingly, a change in the shape of large particles from circular to elliptical resulted in selective entrapments of smaller circular particles, but enhanced outflows of larger particles, which could be an alternative microfluidics-based method for sorting and separation of particles of different sizes and shapes.

\vspace {0.2 cm}

\noindent \textbf{Remarks:}  Interparticle steric interactions can be computed between the boundary nodes on the solid surfaces of particles in near contact, rather than relying on their center-to-center separation distance. This makes the model readily applicable to both curve-shaped (e.g., elliptical, circular) and angular-shaped (e.g., boomerang-shaped, star-shaped) particles. If the steric interactions are used to avoid overlapping, the stiffness parameter is chosen such that the repulsive force is equal or slightly larger than the hydrodynamic forces until the separation distance exceeds the threshold interaction distance. For the particle entrapment and channel clogging problems, the stiffness parameter and the surface asperities are the vital parameters. Larger repulsive forces than hydrodynamic forces could lead to temporary entrapment of a particle in a flow channel. A comparison between LJ models is demonstrated in Table \ref{LJ_Comp}.

  %%%%%%%%%%%%%%%%%%%%%%%%%%%%%%%%%%%%%%%%%%%%%%%%%%%%%%%%%%%%%%%%%%%%%%%%%%%%%%%%%%%%%%%%%
%---------------------------------------------------------------------
\begin{table}[! htbp]
\centering 
\caption{A comparison of different Lennard-Jones potential models.}
\small
\tabcolsep=0.11cm
\begin{threeparttable}
\begin{tabular}{l c c c c c c}
\toprule
\thead{ Model  } & \thead{Method} 
 &\thead{Experimental \\ Validation}
 &\thead{Particle \\ Shape}
 &\thead{Parameters}
 &\thead{Normal \\ Force}
 &\thead{Tangential \\ Force} \\ \midrule
    Model 1  &  \thead{IB-LBM}         & No  & Circular  & $\zeta, \triangle x, \varepsilon, R_a, R_b$ &  Yes  & No   \\
    Model 2  &  \thead{Colloidal\\LBM} & No & \thead{ Circular or \\ non-Circular}  & $\zeta, \triangle x, \varepsilon$ &  Yes  & Yes\tnote{*}   \\

\bottomrule\addlinespace[1ex]
\end{tabular}
\begin{tablenotes}\footnotesize
\item[*] Tangential force naturally emerges due to the consideration of repulsive interaction forces among the boundary nodes of neighboring particles.  
\end{tablenotes}
\end{threeparttable}
\label{LJ_Comp}
\end{table}
%---------------------------------------------------------------------

%%%%%%%%%%%%%%%%%%%%%%%%%%%%%%%%%%%%%%%%%%%%%%%%%%%%%%%%%%%%%%%%%%%%%%%%%%%%%%%%%%%%%
\section{Collision Forces through the Discrete Element Modeling}
\label{sec4}
%%%%%%%%%%%%%%%%%%%%%%%%%%%%%%%%%%%%%%%%%%%%%%%%%%%%%%%%%%%%%%%%%%%%%%%%%%%%%%%%%%%%% 

\begin{figure}[ht!]
\centering
\includegraphics[width=0.9\textwidth]{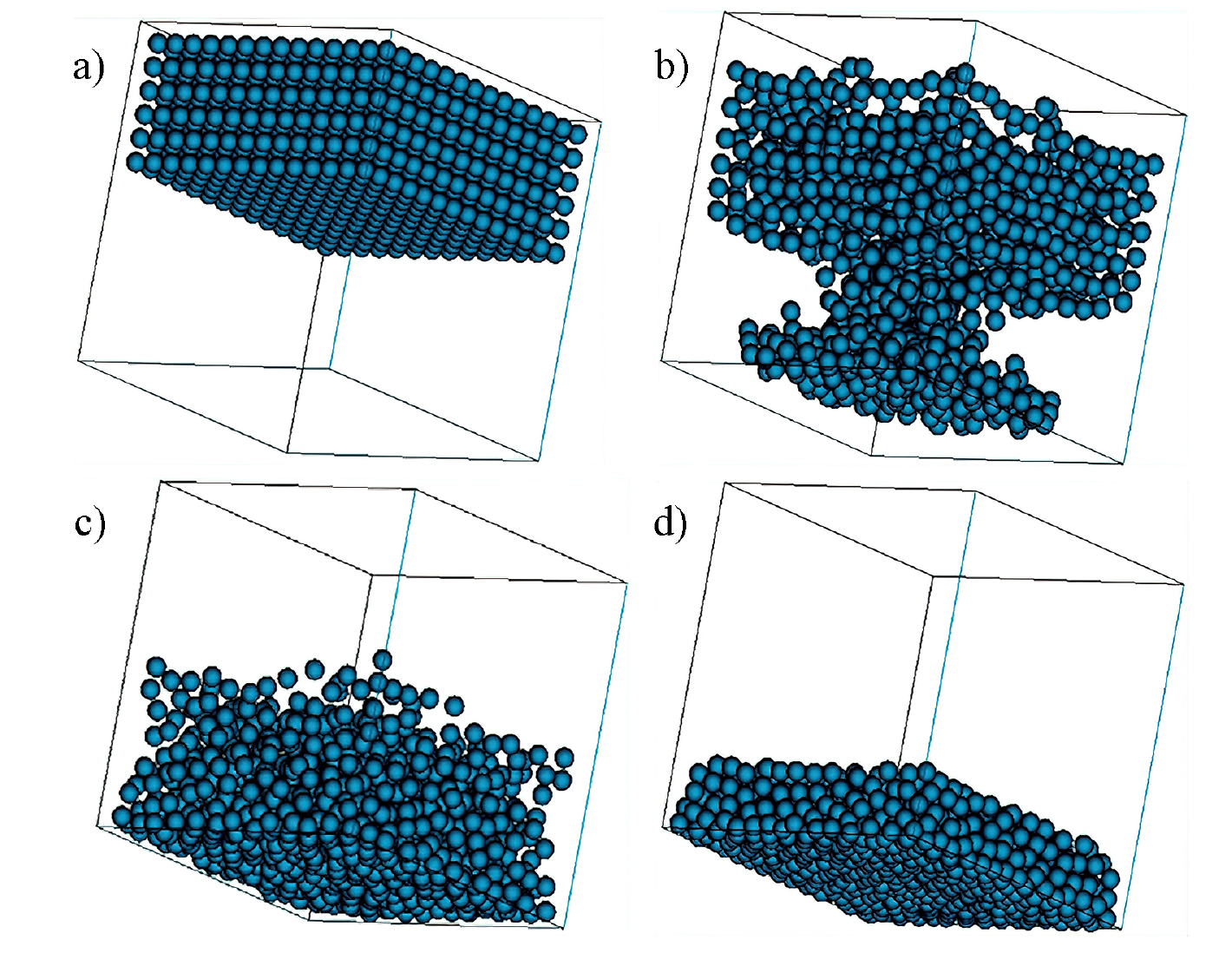}
\caption{IBM-LBM-SPM simulation of settling of 1,176 particles in a bounded domain at (a) $t$=0.0 s, (b) $t$=5.7 s, (c) $t$=11 s, and (d) $t$=20 s (adapted from \cite{khalili2017} under conditions of the Creative Commons Attribution CC BY license). }
\label{fig:Fig13}
\end{figure}

Zhang et al. \cite{ZYTY_CMwA_16} used IBM-LBM to simulate the sedimentation of 8,125 spherical particles, in which they computed collision forces through DEM when the particles collide with each other or with the walls. The authors described the particles and walls by material properties such as density, Young’s modulus ($E_i$), Poisson’s ratio ($\nu_i$) and friction coefficient. For a spherical particle of radius $R_i$, they computed the normal force–displacement relationship between the colliding particles by

\begin{equation}
 {F}_{n}= \frac{4}{3} E^\ast \left({R^\ast}\right)^{0.5} \delta_n^{1.5},
 \label{e55}
 \end{equation}

\noindent where $1/E^\ast=(1-\nu_1^2 )+(1-\nu_2^2 )/E_1$ and $1/R^\ast= 1/R_1+1/R_2$. The incremental tangential force resulting from an incremental tangential displacement depends on the loading history and the normal force 

\begin{equation}
 \triangle T= 8 G^\ast r_a \theta_k \triangle \delta_t + \left( -1 \right)^k \Upsilon \triangle F_n \left(1-\theta_k \right) ,
 \label{e56}
 \end{equation}

\noindent where $1/G^\ast= (1-\nu_1^2 ) G_1 + (1-\nu_2^2 )/G_2$,  $R_a= \sqrt{\left(\delta_n R^\ast \right)}$ is the radius of the contact area, $\delta_t$ is the relative tangential incremental surface displacement, $\Upsilon$ is the coefficient of friction, the value of $k$ and $\theta_k$ change with the loading history. Using this collision force calculation scheme, the authors simulated the settling of 8,125 spherical particles. Similarly, Khalili et al. \cite{khalili2017} simulated the particulate suspensions by the LBM combined with a smoothed-profile method (SPM) in an IBM-LBM-SPM scheme. In the latter case, an artificial repelling force was added to prevent spheres from penetrating each other or with the wall. Finally, the extra short-range repulsive force was taken with a quadratic parabolic functional form on the distance among particles, showing the success in simulating the sedimentation of 1,176 lagrangian spheres in a quiescent fluid (see Figure \ref{fig:Fig13}).

%%%%%%%%%%%%%%%%%%%%%%%%%%%%%%%%%%%%%%%%%%%%%%%%%%%%%%%%%%%%%%%%%%%%%%%%%%%%%%%%%%%%%%%%%%
%----- Figure 12 ----
%%%%%%%%%%%%%%%%%%%%%%%%%%%%%%%%%%%%%%%%%%%%%%%%%%%%%%%%%%%%%%%%%%%

%%%%%%%%%%%%%%%%%%%%%%%%%%%%%%%%%%%%%%%%%%%%%%%%%%%%%%%%%%%%%%%%%%%%%%%%%%%%%%%%%%%%%%%%%%

%%%%%%%%%%%%%%%%%%%%%%%%%%%%%%%%%%%%%%%%%%%%%%%%%%%%%%%%%%%%%%%%%%%%%%%%%%%%%%%%%%%%%%%%%%

\vspace {0.2 cm}

\noindent \textbf{Remarks:} Collision forces are computed based on tangential and normal displacements. The model was successfully used with a large number of spherical particles in isothermal and thermal conditions. Its further extension to a large number of mixtures of non-spherical particles in Newtonian or non-Newtonian fluid flows could make the upgraded model a versatile numerical tool for particle flow simulations.

 %%%%%%%%%%%%%%%%%%%%%%%%%%%%%%%%%%%%%%%%%%%%%%%%%%%%%%%%%%%%%%%%%%%%%%%%%%%%%%%%%%%%%
\section{Conclusions and Outlook}
\label{Sec5}
%%%%%%%%%%%%%%%%%%%%%%%%%%%%%%%%%%%%%%%%%%%%%%%%%%%%%%%%%%%%%%%%%%%%%%%%%%%%%%%%%%%%% 

Close-contact interactions, hydrodynamics, and collisions are crucial to the sedimentation and flow of non-uniform-shaped and -sized particles in fluidic systems. At the benchmark level, numerical and experimental studies have been performed to quantify near-contact and direct-contact dynamics of two particles as they settle in an initially quiescent fluid, known as the Drafting, Kissing and Tumbling (DKT) problem. These analyses unveiled valuable insights into the effect of the size, shape, temperature, and initial position of the particles, and the heterogeneous, thermal, non-Newtonian, and viscoelastic nature of the host fluid on hydrodynamic interactions of the particles. The DKT analyses, however, have been typically performed with curved-shaped particles. These analyses should be extended to sedimentation of a mixture of angular- and curved-shaped particles, as shown in Fig. 7, as surfaces of the particles are often not curved. In-depth quantitative understandings of the effect of particle shape on their close-contact dynamics and collisions are vital to quantify macroscale particles’ hydrodynamics accurately.  

Near-contact and direct-contact hydrodynamics of the particles have been simulated using mathematically and physically simple to complex short-range interaction models, such as  lubrication force (LF), spring-force (SF), soft-sphere (SSM), and hard-sphere models (HSM), and/or combinations thereof. These models have been coupled to the IB-LBM to simulate the settling or flow of deformable or rigid particles. From a mathematical standpoint, the distinction between some of the LF and SF models is somehow hazy. Most of the LF models in the literature require the presence of at least one fluid node between adjacent solid surfaces of mobile bodies to resolve hydrodynamics of the thin fluid film between the surfaces in near-contact. However, alternative LF that eliminates the need for a fluid node between solid surfaces near contact or a stiffness parameter has been proposed. Some of the LF and SF models allow particles to overlap, and the overlapped distances are used to calculate particle-particle collision forces. The LF and SF model can be used to avoid particles collision by introducing short-range surface-to-surface steric interactions or by imposing post-collision repulsion by calculating short-range repulsive forces based on the collision forces. Most of these repulsive models require a stiffness parameter, measuring the strength of the repulsive interactions, and/or the threshold interaction distance beyond which the repulsive interactions vanish. Although these parameters are often defined as a function of the computational grid spacing, there is no consolidated theory or a universally accepted method to determine their optimal values. As a result, conflicting results have been reported in the literature on the importance of short-range interparticle and particle-wall repulsive interactions on the sedimentation and flow dynamics of the particles in simulations. Various short-range repulsive models reviewed in this paper fall short of accommodating the effect of different sizes and/or shape of the particles, local flow field, or local viscosity. Therefore, we envision that an ideal short-range repulsive model should accommodate the effect of the fluid velocity, local viscosity, and close contact and collision dynamics of a mixture of arbitrary-shaped particles. The model should also allow the particles to overlap to enhance the numerical stability and calculate repulsive forces based on their collision dynamics, overlapping distance, and interaction forces. Alternatively, SSM and HSM have been used to simulate close contact and direct contact dynamics of the particles. If the number of particles is small, the HSM could be computationally more efficient than the SSM, as the contact time in the soft-sphere model is typically much longer than the time-step of the simulations. The SSM, however, can handle multiple interparticle contacts and overlapped surfaces.   

The Lennard-Jones (LJ) 6-12 potential model has also been used to simulate close contact dynamics of the particles. Although one of the LJ models reviewed herein is free of calibration parameters, it calculates the short-range repulsive interaction forces as a function of the center-to-center distance between the adjacent spherical particle in close contact, therefore it is not suitable to account for the effect of surface heterogeneities in close contact hydrodynamics. The alternative LJ model reviewed in this paper calculates the interaction forces between the boundary nodes of two particles in near contact; hence, it can effectively accommodate the effect of surface heterogeneities and the tangential component of the steric forces in interaction force calculations. However, it relies on problem-specific stiffness parameters and the threshold interaction distance. The latter version was successfully used for the arbitrary-shaped particles in Newtonian or non-Newtonian fluidic environments. Therefore, the ideal LJ model would be the combination of these two LJ models. Potential coupling of the LJ model with the SSM or collision force models appears to be a promising yet unexplored research field. Moreover, the existing particles collision models need further upgrades to simulate close-contact and collision dynamics of a mixture of arbitrary-shaped particles in Newtonian or non-Newtonian fluid flows to enhance their uses in practice. Thus, there is more scope for improvements in close-contact and direct-contact force calculations, which are imperative for more accurate calculations of multi-body interactions among a mixture of settling or flowing nonuniform-sized and -shaped particles in diverse natural and industrial complex flows. 

%%%%%%%%%%%%%%%%%%%%%%%%%%%%%%%%%%%%%%%%%%%%%%%%%%%%%%%%%%%%%%%%%%%%% 

%%%% Acknowledgments %%%%%%%%
\section*{Acknowledgments}
G.F. acknowledges the support of PRIN projects CUP E82F16003010006 (principal investigator, G.F. for the Tor Vergata Research Unit) and CUP E84I19001020006 (principal investigator, G. Bella). S.S. acknowledges financial support from the European Research Council under the Horizon 2020 Programme advanced grant agreement no. 739964 (‘COPMAT’).\\

%%%% Bibliography  %%%%%%%%%%

\end{document}